\documentclass[10pt,conference,letterpaper]{IEEEtran}
\usepackage{times,amsmath,epsfig}

\usepackage{booktabs} 

\usepackage{esvect}
\usepackage{etoolbox}
\usepackage{graphicx}
\usepackage{balance}  
\usepackage{todonotes}
\usepackage{subcaption}
\usepackage{url}
\usepackage{cite}

\usepackage{setspace}

\newcommand\ceil[1]{\lceil#1\rceil}

\usepackage{listings}
\usepackage[framemethod=TikZ]{mdframed}
\usepackage{xcolor}
\definecolor{mygreen}{rgb}{0,0.6,0}
\definecolor{myblue}{rgb}{0,0,0.8}
\definecolor{mygray}{rgb}{0.6,0.6,0.6}
\definecolor{mypurple}{HTML}{9900FF}
\definecolor{redmaterial}{HTML}{F44336}

\lstdefinestyle{simd}{
  language=C++,
  basicstyle=\small\sffamily,      
  breakatwhitespace=false,         
  breaklines=true,                 
  commentstyle=\color{mygreen},    
  deletekeywords={not, or, and, using},            
  escapeinside={(*}{*)},           
  extendedchars=true,              
  keepspaces=true,                 
  keywordstyle=\color{myblue},     
  numbersep=5pt,                   
  numberstyle=\tiny\color{mygray}, 
  rulecolor=\color{black},         
  showspaces=false,                
  showstringspaces=false,          
  showtabs=false,                  
  stepnumber=1,                    
  tabsize=1,                       
  moredelim=[is][\color{red}]{@}{@},
  moredelim=[is][\color{mypurple}]{$}{$},
  morekeywords={each, __m512i, __mmask8, uint8_t, uint16_t, uint32_t, uint64_t, int8_t, int16_t, int32_t, int64_t}
}

\usepackage{etoolbox}{
\mdfdefinestyle{listingstyle}{%
  backgroundcolor=black!15,outerlinewidth=0.25pt,outerlinecolor=black,%
  innerleftmargin=5pt,innerrightmargin=5pt,innertopmargin=0pt,innerbottommargin=0pt%
}

\def\ParHead{\vspace*{2mm}\noindent\bf}

\usetikzlibrary{arrows.meta}
\tikzset{square arrow/.style={to path={-- ++(0,-.25) -| (\tikztotarget)}}}
\tikzset{square arrow 0/.style={to path={-- ++(0,-.15) -| (\tikztotarget)}}}
\tikzset{square arrow 1/.style={to path={-- ++(0,-.25) -| (\tikztotarget)}}}
\tikzset{square arrow 2/.style={to path={-- ++(0,-.35) -| (\tikztotarget)}}}
\tikzset{square arrow 3/.style={to path={-- ++(0,-.45) -| (\tikztotarget)}}}
\tikzset{square arrow top/.style={to path={-- ++(0,+.25) -| (\tikztotarget)}}}
\tikzset{square arrow top 0/.style={to path={-- ++(0,+.15) -| (\tikztotarget)}}}
\tikzset{square arrow top 1/.style={to path={-- ++(0,+.25) -| (\tikztotarget)}}}
\tikzset{square arrow top 2/.style={to path={-- ++(0,+.35) -| (\tikztotarget)}}}
\tikzset{square arrow top 3/.style={to path={-- ++(0,+.45) -| (\tikztotarget)}}}

\title{Adaptive Geospatial Joins for Modern Hardware}

\author{
\IEEEauthorblockN{Andreas Kipf\IEEEauthorrefmark{1},
Harald Lang\IEEEauthorrefmark{1},
Varun Pandey\IEEEauthorrefmark{1},
Raul Alexandru Persa\IEEEauthorrefmark{1},\\
Peter Boncz\IEEEauthorrefmark{2},
Thomas Neumann\IEEEauthorrefmark{1},
Alfons Kemper\IEEEauthorrefmark{1}}

\vspace{1.6mm}\\

\begin{minipage}{0.5\linewidth}
\centering
\IEEEauthorrefmark{1}Technical University of Munich\\
Munich, Germany\\
\{first.last\}@in.tum.de
\end{minipage}

\begin{minipage}{0.5\linewidth}
\centering
\IEEEauthorrefmark{2}Centrum Wiskunde \& Informatica\\
Amsterdam, The Netherlands\\
boncz@cwi.nl
\end{minipage}

\vspace{3.2mm}\\
}

\begin{document}

\newcommand\codevec[1]{\vv{\text{#1}}}
\newcommand\codevar[1]{\textrm{\textit{#1}}}
\newcommand*\ric[1]{\vphantom{#1}\smash{#1_{}\kern-\scriptspace}}

\lstMakeShortInline[style=simd,columns=fullflexible]|

\maketitle

\begin{abstract}
Geospatial joins are a core building block of connected mobility applications.
An especially challenging problem are joins between streaming points and
static polygons.
Since points are not known beforehand, they cannot be indexed.
Nevertheless, points need to be mapped to polygons with low latencies to enable
real-time feedback.

We present an adaptive geospatial join that uses true hit filtering to avoid
expensive geometric computations in most cases.
Our technique uses a quadtree-based hierarchical grid to approximate polygons
and stores these approximations in a specialized radix tree.
We emphasize on an approximate version of our algorithm that guarantees a
user-defined precision.
The exact version of our algorithm can adapt to the expected point distribution
by refining the index.
We optimized our implementation for modern hardware architectures with wide
SIMD vector processing units, including Intel's brand new Knights Landing.
Overall, our approach can perform up to two orders of magnitude faster than
existing techniques.
\end{abstract}

\section{Introduction}
Connected mobility companies need to process vast amounts of location data in
near real-time to run their businesses.
For instance, Uber needs to join locations of passenger requests with a set of
predefined polygons to display available products (e.g., Uber X) and to enable
dynamic pricing\footnote{\url{https://eng.uber.com/go-geofence/}}.
Another example are traffic use cases where the positions of vehicles need to
be joined with street segments to enable real-time traffic control.
With a future of connected (and possibly self-driving) cars and drones,
high-performance geospatial joins will become a key feature for increasingly
demanding applications.

While geospatial joins have been studied for decades, many of them optimize for
I/O operations, which is an important performance factor for disk-based systems.
With the large main-memory capacities of modern hardware, however, it is for
the first time possible to maintain fine-grained index structures purely in
main memory.

We propose an adaptive geospatial join that addresses workloads with streaming
points and static polygons.
Our approach is optimized for modern hardware with large high-bandwidth
memory and our implementation fully utilizes many-core processors and their
Single Instruction Multiple Data (SIMD) units.

\begin{figure}
\centering
\includegraphics[width=0.65\linewidth]{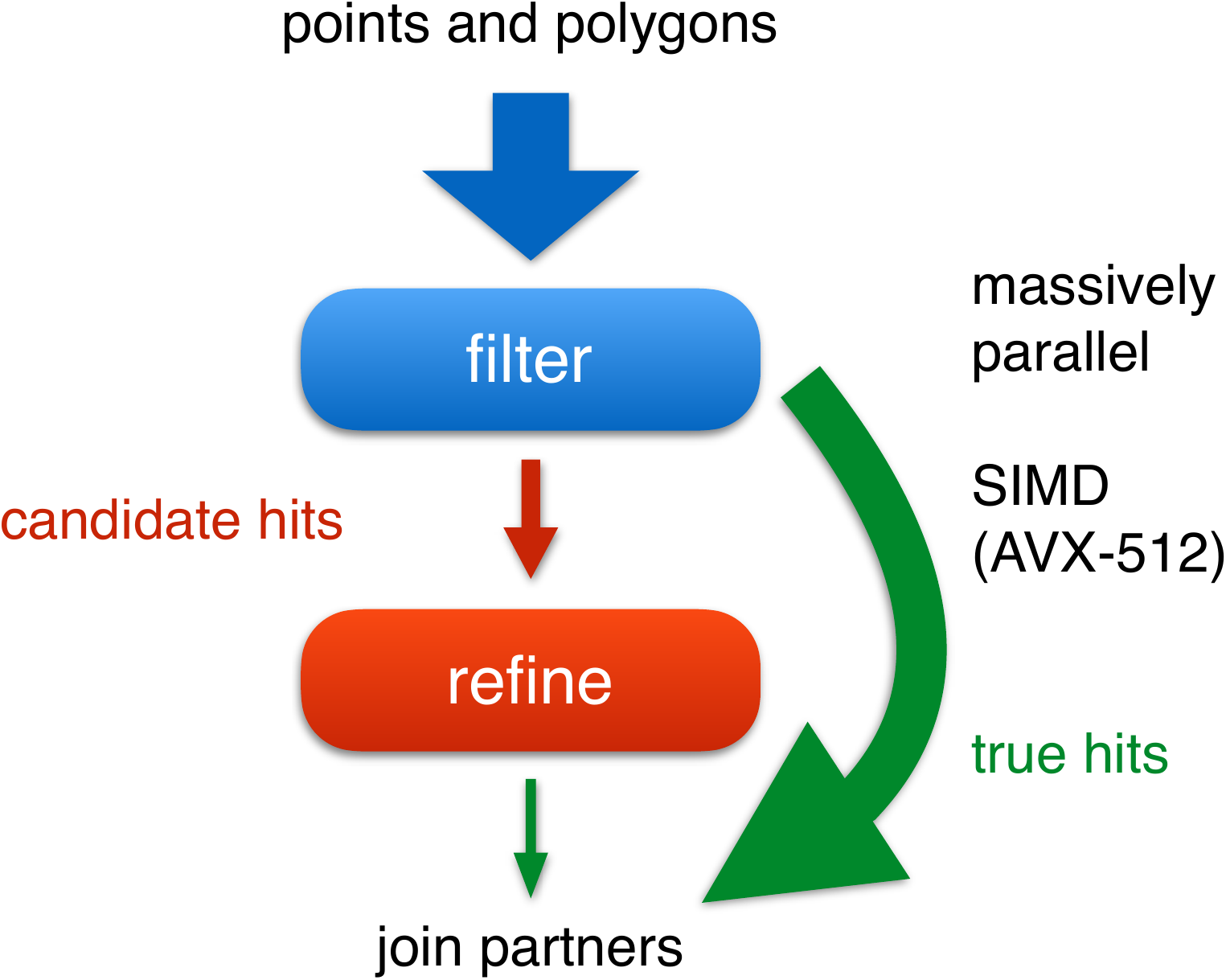}
\caption{We adaptively compute a fine-grained (true hit) filter that identifies
most or even all join partners. We can treat candidate hits as true hits while
guaranteeing a user-defined precision. The arrow width reflects the amount of
data passing through each of the respective paths.}
\label{fig:filter_and_refine}
\vspace{-1em}
\end{figure}

Figure~\ref{fig:filter_and_refine} illustrates our approach.
We make use of \emph{true hit filtering} proposed by~\cite{brinkhoff1994multi}.
The authors of this work used maximum enclosed rectangles and circles as
progressive approximations of single polygons.
In contrast, we use an adaptive grid represented by a specialized radix tree,
Adaptive Cell Trie (ACT), to not only index single polygons but entire sets of
polygons.

There are systems (e.g., Oracle Spatial~\cite{kothuri2001efficient} and Spark Magellan\footnote{\url{https://github.com/harsha2010/magellan}}) that have used
the idea of true hit filtering.
We take this idea one step further and leverage the large main-memory capacities
available in modern hardware to maximize the likelihood of true hits.
We increase this likelihood by training the index using historical data points.

In summary, we transform the traditionally compute-intensive problem of
geospatial joins into a memory-intensive one.
Our approach results in speedups of up to two orders of magnitude compared
to state-of-the-art standalone libraries and geospatial database systems.

Our contributions include:
\begin{itemize}
  \item An adaptive geospatial join that uses hierarchical grids combined with
  radix indexing techniques to identify most join partners in the filter phase.
  \item An approximate version of this approach that guarantees a user-defined
  precision.
  \item An optimization of this technique for Intel's latest Xeon Phi, named
  Knights Landing (KNL).
\end{itemize}

\section{Background}\label{sec:background}
Unless stated otherwise, we follow the semantics of the |ST_Covers| join predicate supported by PostGIS and Oracle Spatial.
|ST_Covers| evaluates whether one geospatial object (e.g., a polygon) covers another (e.g., a point).
Points on the edges or vertices of a polygon are considered to be within the polygon.
Other predicates, such as |ST_Intersects|, are not explicitly addressed in this paper but can be implemented with similar techniques.

Our geospatial join uses several building blocks:

{\ParHead PIP:}
A point-in-polygon~(PIP) test determines whether a point lies within a polygon.
Typically such a test is performed using complex geometric operations, such as the \emph{ray tracing algorithm} which involves drawing a line from the query point to a point that is known to be outside of the polygon and counting the number of edges that the line crosses.
If the line crosses an odd number of edges, the query point lies within the polygon.
The runtime complexity of this algorithm is $O(n)$ with $n$ being the number of edges.
While there are many conceptual optimizations to the PIP test, this operation remains computationally expensive
since it processes real numbers (e.g., latitude/longitude coordinates) and thus involves floating point arithmetics.
In geospatial joins, this test should thus be avoided whenever possible.

{\ParHead Space-filling curves:}
Geographical coordinates can either be indexed using multi-dimensional
data structures, such as R-trees, or they can be linearized and indexed using
one-dimensional data structures, such as B-trees or tries.
Linearization methods include the Hilbert space-filling curve and the Z curve.
Our approach relies on discretization and linearization but does not depend
on a concrete space-filling curve.
For our approach to work, the cell enumeration must only fulfill the property
that children cells share a common prefix with their parent cell.

{\ParHead Google S2:}
We use the Google S2 library\footnote{\url{http://code.google.com/archive/p/s2-geometry-library/}}
that provides primitives that our approach builds upon.
These primitives include a PIP test algorithm and adaptive grid approximations
of polygons.
In general, S2 offers many features for processing geospatial data in both discrete and non-discrete space.
To transform coordinates from the non-discrete space (the latitude/longitude coordinate system) into the discrete space, S2 maps points on earth onto a surrounding unit cube.
Then it recursively subdivides each of the six faces of the cube in a quadtree fashion and enumerates the cells using the Hilbert space-filling curve.
The enumerated cells are 64 bit integers, called \emph{cell ids}, that uniquely identify a cell.
The three most significant bits represent one of the six faces of the cube.
The following zero to 30 bit pairs identify the quadtree cell.
The next bit is always set and is used to identify the level of a cell.
The remaining bits (if any) are set to zero.
The smallest cells use all 64 bits.

\begin{figure}[t!]
\centering
\includegraphics[width=\linewidth]{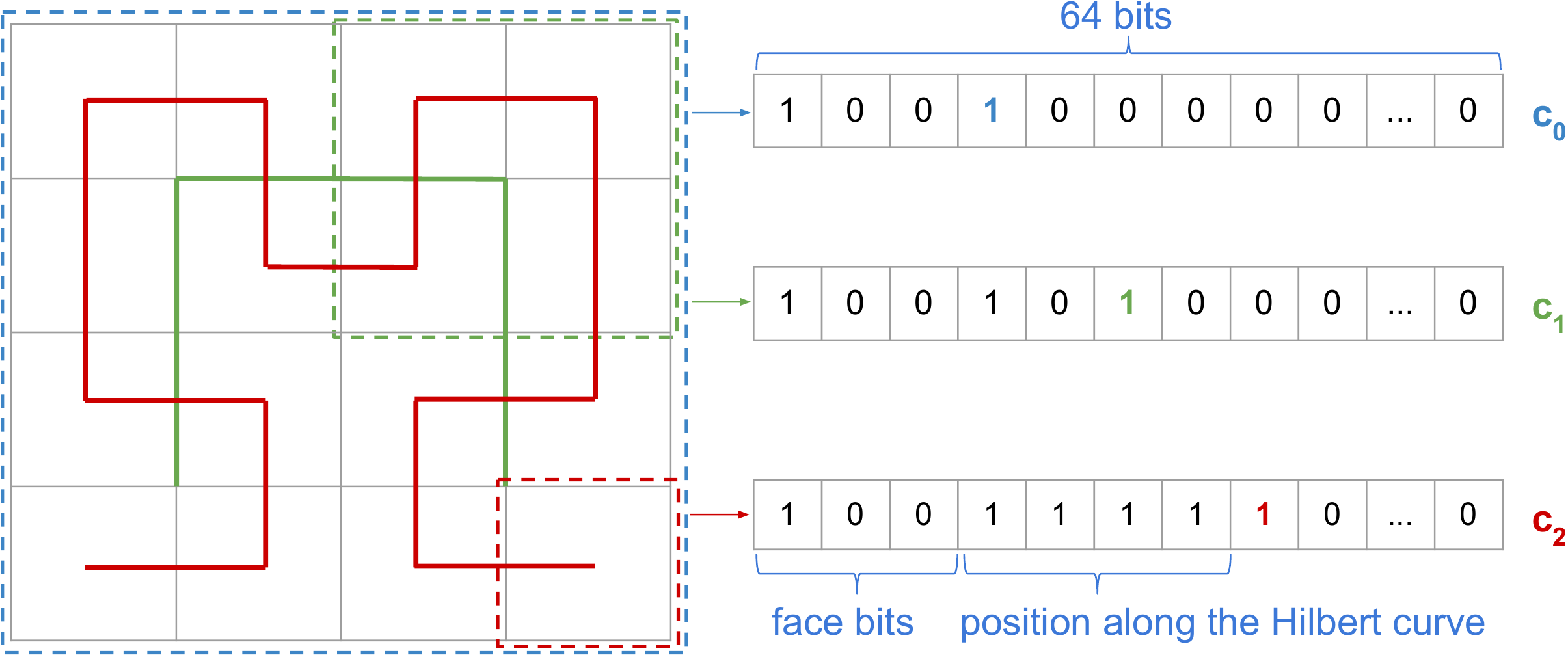}
\caption{Cells at three different levels enumerated by the Hilbert curve and their bit representations}
\label{fig:hilbert}
\end{figure}

Figure~\ref{fig:hilbert} illustrates the hierarchical decomposition of a cube face
and shows how many bits the corresponding cell ids require depending on their
level.
The colored bits encode the levels.
The cell ids in the example share a common prefix with their ancestors up
to the colored bit.
For example, $c_0$ is an ancestor of $c_1$ and $c_2$ (i.e., $c_0$ fully contains $c_1$ and $c_2$).
A containment relationship between cells can thus be efficiently computed using
bitwise operations.

\begin{figure}[t!]
  \centering

  \begin{subfigure}[b]{0.42\linewidth}
    \includegraphics[width=\linewidth]{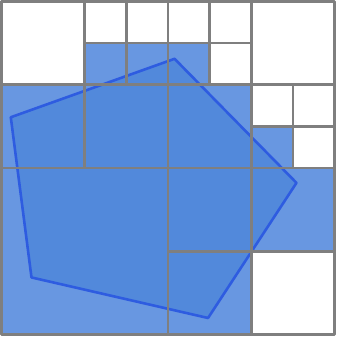}
    \caption{Covering}
    \label{fig:covering}
  \end{subfigure}
  ~
  \begin{subfigure}[b]{0.42\linewidth}
    \includegraphics[width=\linewidth]{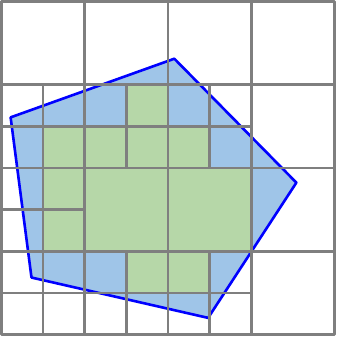}
    \caption{Interior covering}
    \label{fig:interior_covering}
  \end{subfigure}
  \caption{A covering and an interior covering of a polygon}
  \label{fig:coverings}
  \vspace{-1em}
\end{figure}

As mentioned above, S2 offers methods for approximating polygons.
Figure~\ref{fig:covering} shows a covering of a polygon.
A covering is a collection of cells (possibly at different levels) fully covering a polygon.
Figure~\ref{fig:interior_covering} shows an interior covering.
As the name suggests, an interior covering approximates the interior of a polygon.
All cells of an interior covering (interior cells) are completely contained within the polygon.
When a point lies in an interior cell, we know that it also lies within the polygon.
Both types of coverings can be utilized to speed up PIP tests.

To allow for an efficient search, S2 stores the cell ids of a covering in a sorted vector.
Besides sorting the cell id vector, it allows for \emph{normalizing} the covering.
A normalized covering neither contains conflicting nor duplicate cells.
A conflict between two cells exists, if one cell contains the other.
Only when the covering is normalized, cell containment checks can be efficiently ($O(log\,n)$)
implemented using a binary search on the sorted vector.
The normalization of an covering does not lead to a precision loss.

\section{Geospatial Join}\label{sec:approach}
With our approach, we address geospatial joins between streaming points and static
polygons.
An example of such a workload is mapping Uber or DriveNow cars and passenger
requests to predefined zones for allocation purposes.

\subsection{Overview}
Our geospatial join takes a set of simple polygons, builds an index, and probes
points against the index.
The high level idea is to index adaptive grid approximations of the polygons
in a highly-efficient and specialized radix tree.
In contrast to techniques that first reduce the number of polygons using an index,
e.g., an R-tree on the polygons' minimum bounding rectangles (MBRs),
and then refine candidates using geometric operations,
our approach implements true hit filtering~\cite{brinkhoff1994multi}
and identifies most or even all join partners in the filter phase.
Given a memory budget and a precision bound, we adapt our index to satisfy
these constraints while maximizing probe performance.

Our algorithm consists of five phases:

{\ParHead Build logical index:} We compute coverings and interior coverings of
all polygons and merge them to form a \emph{logical} index.

{\ParHead Build physical index:} Then we build a \emph{physical} index on the cells of the logical index.

{\ParHead Training:} When our approximate approach cannot guarantee a
user-defined precision without exceeding a specified memory budget, we use an
exact approach and train the index with historical data points to increase
probe performance.
Popular areas that expect more hits are approximated using a finer-grained grid
than less popular areas.

{\ParHead Probe index:} In this phase, we probe the points against the physical index.
An index probe can either return a \emph{false hit} or a list of \emph{hits},
where each \emph{hit} can either be a \emph{true hit} or a \emph{candidate hit}.

{\ParHead Refine candidates:} Finally, we perform exact geometric computations
for all candidate hits\footnote{Often, this phase can be skipped. In fact, it is only required if exact results are needed and if there are candidate hits at all.}.
\\

We use the S2 library to compute coverings and interior coverings.
Note that our technique does not rely on S2.
In fact, any other adaptive grid index would work as well.
We then merge the individual coverings to create a logical index,
before we build a physical index.
Note that the first two phases can be performed in one step,
i.e., the individual coverings can be computed and merged while inserting their cells into the index.
In our approach, the probe (filter) phase is the performance critical part.
We therefore highly parallelize this phase using thread- and instruction-level
parallelism to accelerate lookups in the physical index.
In the refinement phase, we use S2's PIP test, which implements the
ray tracing algorithm (cf.,~\cite{pandey2016high} for performance numbers).

The overall strategy of our technique is to minimize the number of (expensive)
refinements.
We can decrease the number of refinements by using a more fine-grained index.
Since we make use of true hit filtering, a precise index allows us to
identify \emph{most} join partners during the filter phase.
We can always omit the refinement phase when approximate results are sufficient.
The error of such an approximate algorithm is bound by the length of the diagonal
of the largest covering cell, which we choose to be sufficiently small.

Our geospatial join takes two parameters: (i) a memory budget that must not be
exceeded and (ii) an application-dependent precision\footnote{The maximum
distance (in meters) between the two partners of a false positive join pair.}
that must be guaranteed.
We first try using the approximate approach by refining the index until we can
guarantee the user-defined precision bound.
This involves replacing covering cells with their children cells until the
largest covering cell is sufficiently small (i.e., its diagonal is less than the
precision bound).
It might occur that we exhaust the memory budget during this procedure.
In that case, we use the exact approach and train the index until
we have used the entire memory budget.
Training the index reduces the likelihood for expensive geometric operations.
We evaluate these two approaches in Section~\ref{sec:evaluation}.

In summary, we trade memory consumption with precision (approximate approach)
and probe performance (exact approach).
The approximate approach almost involves no computation (except result aggregation)
while the exact approach reduces expensive computations by training the index.
Thus, both strategies favor modern hardware with large main memory capacities
and high memory bandwidths.

\subsection{Build Logical Index}\label{sec:logicalindex}
\begin{figure}[t!]
  \centering

  \begin{subfigure}[b]{0.30\linewidth}
    \includegraphics[width=\linewidth]{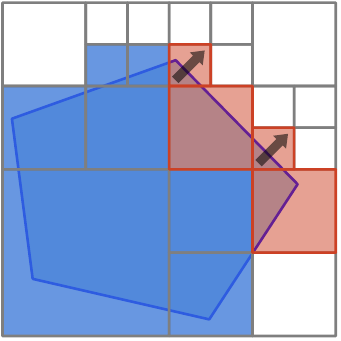}
    \caption{Covering}
    \label{fig:precision_loss_a}
  \end{subfigure}
  ~
  \begin{subfigure}[b]{0.30\linewidth}
    \includegraphics[width=\linewidth]{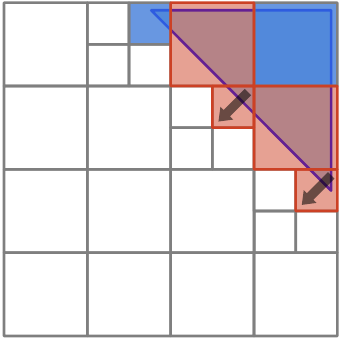}
    \caption{Covering}
    \label{fig:precision_loss_b}
  \end{subfigure}
  ~
  \begin{subfigure}[b]{0.30\linewidth}
    \includegraphics[width=\linewidth]{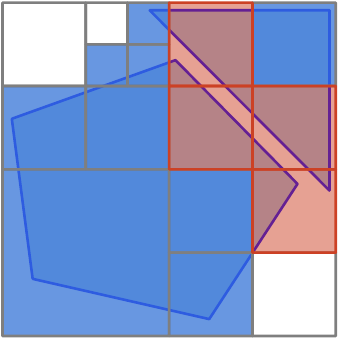}
    \caption{Super cov.}
    \label{fig:precision_loss_c}
  \end{subfigure}
  \caption{A normalized super covering may be less selective than two individual coverings.
  The arrows indicate that the cells will be expanded.}
  \label{fig:precision_loss}
\end{figure}

We compute coverings and interior coverings of all polygons, and combine them
into a logical index, which we call \emph{super covering}.
The precision of the super covering defines the selectivity of the index.
As mentioned earlier, a normalized covering excludes conflicting and duplicate cells.
In contrast to the lossless normalization of a single covering, we might experience a precision loss when merging two overlapping coverings and normalizing the resulting super covering\footnote{There is no precision loss if the overlap only consists of duplicate cells.}.
In other words, a normalized super covering may be less selective than two individual coverings.
Figures~\ref{fig:precision_loss_a} and~\ref{fig:precision_loss_b}
show the coverings of two individual polygons.
The red cells have conflicts with cells of the other covering.
Figure~\ref{fig:precision_loss_c} shows the normalized super covering.
The conflicting cells were expanded to larger cells, which lead to a precision loss.

\begin{figure}[t!]
  \centering

  \begin{subfigure}[t]{0.30\linewidth}
    \includegraphics[width=\linewidth]{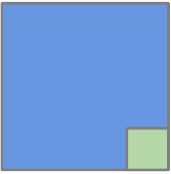}
    \caption{$c_1$ and $c_2$}
    \label{fig:conflict_resolution_a}
  \end{subfigure}
  ~
  \begin{subfigure}[t]{0.30\linewidth}
    \includegraphics[width=\linewidth]{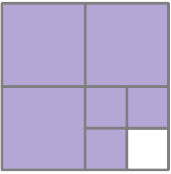}
    \caption{Difference $d$}
    \label{fig:conflict_resolution_b}
  \end{subfigure}
  ~
  \begin{subfigure}[t]{0.30\linewidth}
    \includegraphics[width=\linewidth]{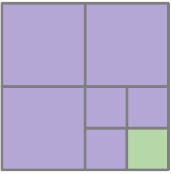}
    \caption{$d$ and $c_2$}
    \label{fig:conflict_resolution_c}
  \end{subfigure}
  \caption{Precision preserving conflict resolution.
  $c_1$ is marked in blue, $c_2$ in green, and the cells in $d$ in purple.
  Note that $c_1$ contains $c_2$.}
  \label{fig:conflict_resolution}
  \vspace{-1.0em}
\end{figure}

When combining the covering and the interior covering of a single polygon, the situation is even more severe since there will be many overlapping cells.
To retain the precision of the individual coverings, we can omit removing conflicting cells and only remove duplicates.
However, there is a trade-off between the size of a super covering and its precision.
We have experimented with three flavors of a super covering:
(i) only covering cells, (ii) covering and interior cells \emph{with} precision loss, and (iii)
covering and interior cells \emph{without} precision loss.
The first approach only indexes covering cells.
This approach cannot identify true hits during the filter phase.
The second approach additionally indexes interior cells and can thus skip the refinement phase when hitting an interior cell.
The first two approaches normalize the super covering when combining the cells
leading to a precision loss.
The third approach avoids the normalization.
Instead of storing a cell $c_1$ and its descendant cell $c_2$, we compute their difference $d$
and store $c_2$ and $d$.
This has the advantage that there will not be any overlap between the indexed cells
and thus an index lookup will at most return a single cell.
The side effect is that the total number of cells will increase since
$d$ consists of at least three cells.
This approach retains the precision and the type (covering or interior) of the individual cells as well as the mappings of cells to polygons\footnote{We expect the individual coverings (that serve as inputs to the merge and index phases) to be normalized. Instead, they could also be normalized ``on the fly''.}.
Figure~\ref{fig:conflict_resolution} illustrates this precision preserving conflict resolution.
Assume that $c_1$ and $c_2$ are cells of two different coverings.
First, we compute $d$, which consists of six cells.
We then copy all references of $c_1$ to $d$ and $c_2$ and omit $c_1$.
Note that the cell count is increased by five.

Each cell in the super covering maps to a list of \emph{polygon references}.
A polygon reference has two attributes:

{\ParHead polygon id:} The id of the polygon that this cell references.

{\ParHead interior flag:} Whether the cell is an interior or a covering cell of
the polygon.

\noindent\begin{figure}
  \begin{mdframed}[outerlinewidth=0.1pt,outerlinecolor=black,
    innerleftmargin=2pt,innerrightmargin=2pt,innertopmargin=-15pt,innerbottommargin=10pt]
{\small{\small
\begin{lstlisting}[caption={Build super covering},label={alg:buildsupercovering},style=simd,columns=fullflexible]
input:
  a list of coverings (*$\codevar{coverings}$*) // one per polygon
  a list of interior coverings (*$\codevar{interiors}$*) // one per polygon
output:
  // a list of (cell, polygon references)
  the super covering (*$\codevar{superCovering}$*)
begin
  for ((*$\codevar{covering}$*) in (*$\codevar{coverings}$*)) {
    for ((*$\codevar{cell}$*) in (*$\codevar{covering}$*)) {
      if ((*$\codevar{superCovering}$*) already contains (*$\codevar{cell}$*)) {
        add references of (*$\codevar{cell}$*) to existing cell
        continue
      }
      if ((*$cell$*) conflicts with existing cell in (*$\codevar{superCovering}$*)) {
        // cell is covered by existing cell or vice versa
        // resolve conflict
        (*$c_1$*) = ascendant cell // may be cell or existing cell
        (*$c_2$*) = descendant cell // may be cell or existing cell
        (*$d$*) = difference of (*$c_1$*) and (*$c_2$*)
        add references of (*$c_1$*) to (*$d$*) and (*$c_2$*)
        remove (*$c_1$*) from (*$\codevar{superCovering}$*) // only required if the existing cell is the ascendant cell
        add (*$c_2$*) and (*$d$*) to (*$\codevar{superCovering}$*)
        continue
      }
      add {(*$\codevar{cell}$*), {(*$\codevar{covering.polygonId}$*), interior flag=false}} to (*$\codevar{superCovering}$*)
    }
  }
  // ... same code for interior coverings (with interior flag=true)
end
\end{lstlisting}}}
\end{mdframed}
\vspace{-1.0em}
\end{figure}

Listing~\ref{alg:buildsupercovering} outlines the algorithm that builds a super covering
of coverings and interior coverings and retains the precision of the individual coverings.
We iterate over all input cells and try to insert them into the super covering.
When a cell already exists, this means that it is also part of another covering that has already been processed.
When a cell conflicts with another cell, this means that either the current cell covers the other cell or vice versa.
These two cases may happen when polygons overlap or are close to each other.
Conflicts also occur when we first insert the cells of a covering of a given polygon and then the cells of its interior covering.
The interior cells always overlap some (if not all) covering cells.
By appropriately resolving these conflicts, we retain the precision of the index.
As mentioned earlier, this strategy increases the total number of cells.
However, a more precise index reduces the number of refinements and thus increases the
overall performance of our algorithm.
In the remainder of this paper, we use the approach that retains the
precision of the individual coverings.

\begin{figure}[t!]
  \centering

  \begin{subfigure}[b]{0.48\linewidth}
    \includegraphics[width=\linewidth]{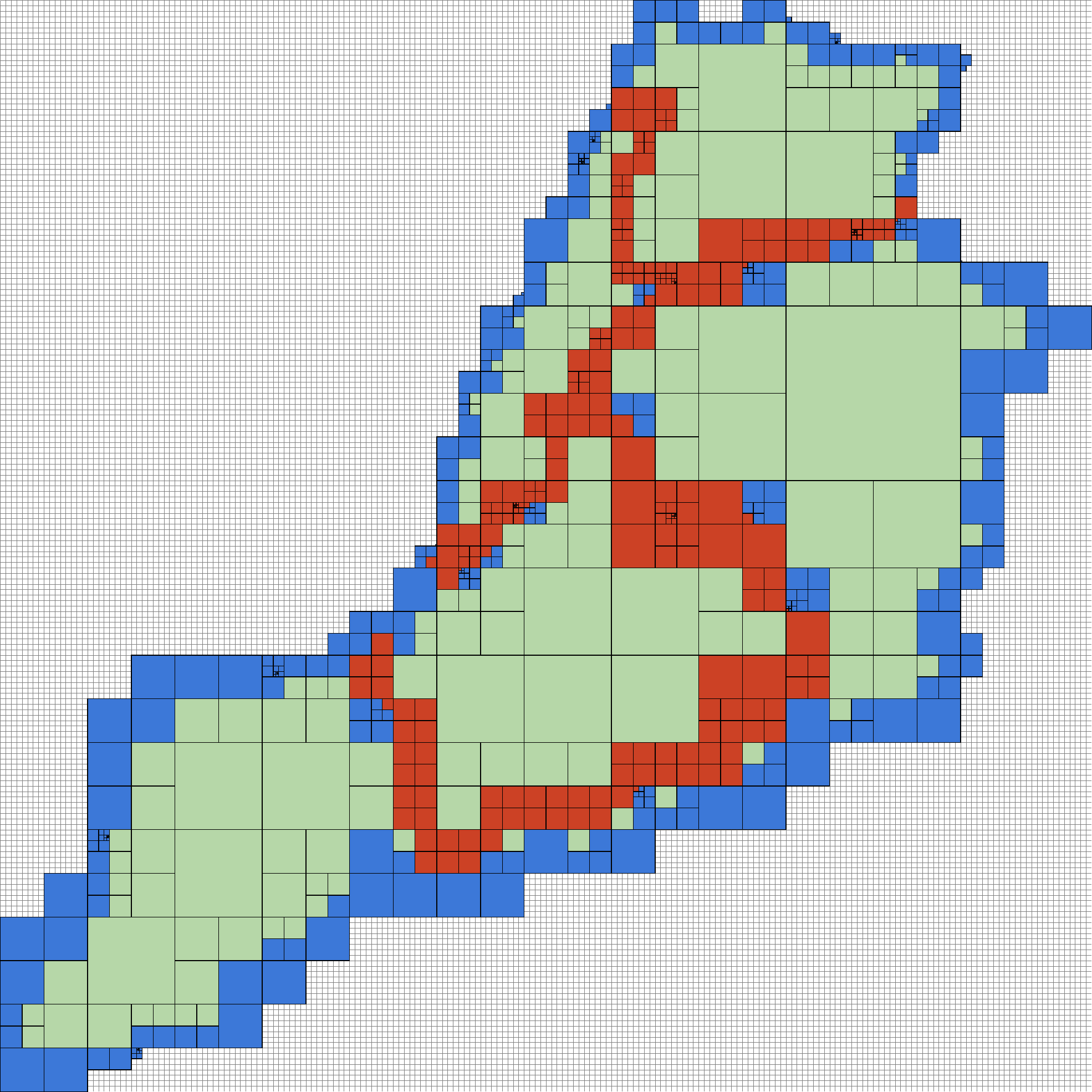}
    \caption{Coarse-grained}
    \label{fig:supercovering_a}
  \end{subfigure}
  ~
  \begin{subfigure}[b]{0.48\linewidth}
    \includegraphics[width=\linewidth]{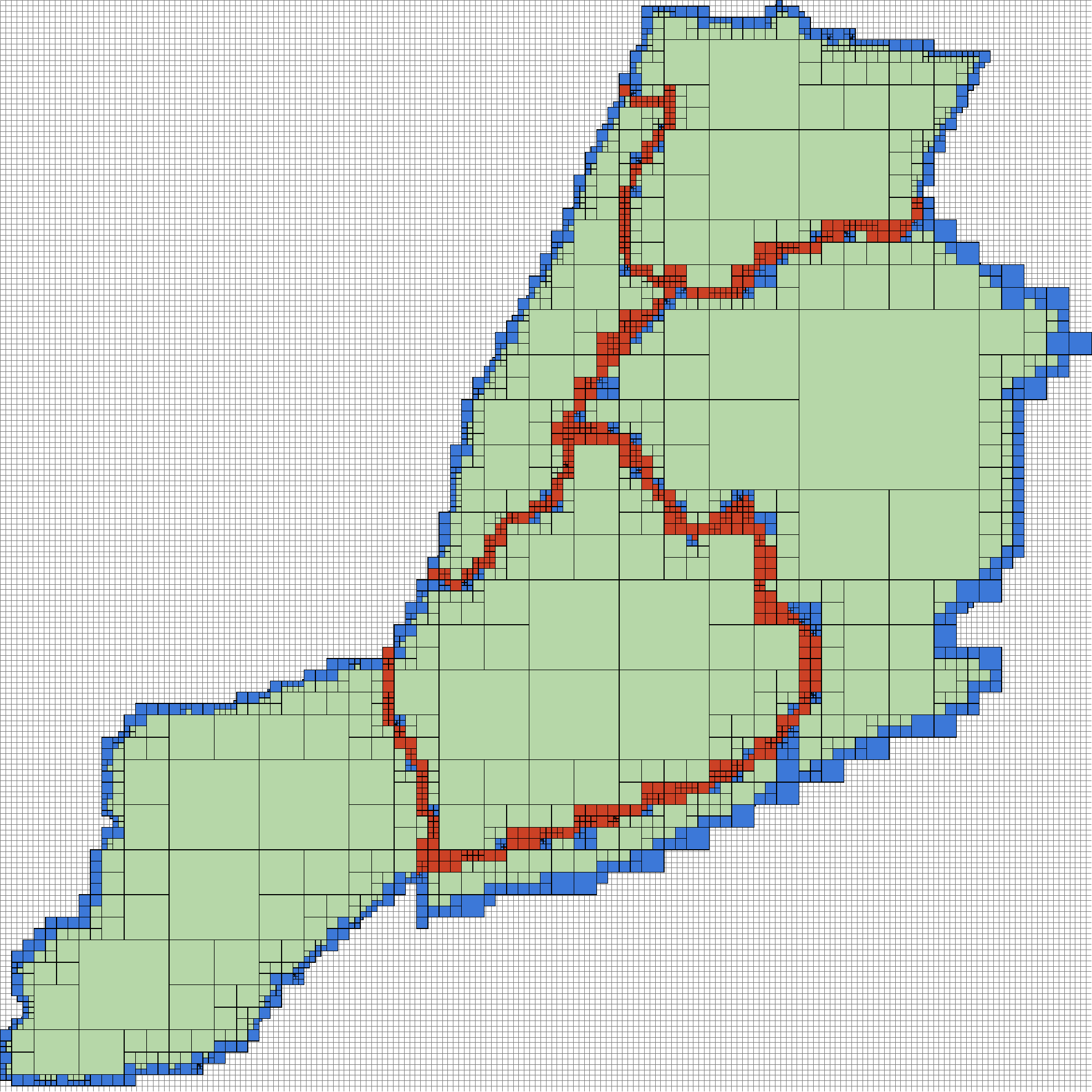}
    \caption{Fine-grained}
    \label{fig:supercovering_c}
  \end{subfigure}
  \caption{Super coverings of the five NYC boroughs.
  Interior cells are marked in green, covering cells of single polygons in blue,
  and covering cells of multiple polygons in red.}
  \label{fig:supercovering}
  \vspace{-1.0em}
\end{figure}

Figure~\ref{fig:supercovering} illustrates a coarse-grained and a fine-grained
super covering of the five NYC boroughs.
For computing the illustrated super coverings, we use up to 50 and up to 200
cells per individual (interior) covering for the coarse-grained and the
fine-grained super covering, respectively.
The key message of this illustration is that the size of the green area
increases with more precise individual coverings, while the size of the red
and the blue area decreases.
Increasing the size of the green area increases the likelihood that we can
skip the refinement phase.
Interior cells are not only more likely to be hit than covering cells
but they also tend to be larger.
Since larger cells use less bits, this leads to an additional performance
improvement when probing ACT.
The reason for this is that large cells are indexed high up in the tree
and are thus found sooner.

\subsection{Build Physical Index}\label{sec:physicalindex}
To store the logical index, we use a physical index
consisting of two data structures, ACT and a lookup table.
Both data structures are designed as in-memory data structures and are
optimized for modern hardware.
In particular, we utilize large main-memory capacities, avoid unnecessary
cache misses, and have parallelized and vectorized the lookups in ACT (cf., Section~\ref{sec:simd}).
ACT stores the cell ids of the super covering and
the lookup table stores the corresponding lists of polygon references.
When a given cell maps to less than three polygon references,
we inline the references into the tree nodes avoiding an additional indirection
and a possible cache miss.

{\ParHead Adaptive Cell Trie:}
In S2, the cells of an individual covering are stored in a sorted vector that can be efficiently searched using a binary search (cf., Section~\ref{sec:background}).
While this approach has a runtime complexity of $O(log\,n)$, it is not very space efficient, since the cell ids of the cells of a single covering often share a common prefix that is stored redundantly.
For a single covering, the space consumption can be reduced by storing the common prefix separately and only storing the varying parts of the cells (e.g., using delta encoding).
Since the cells of a single covering are spatially dense and thus likely to share a common prefix, this works well.
However, when combining the coverings and interior coverings of multiple polygons, the cells are not necessarily dense anymore.
Also, we only require \emph{prefix} lookups, i.e., we need to search for the cell ids that share a common prefix with our search key (the cell id of a query point) to check whether the query point is contained in one of the indexed cells.
A radix tree thus is the ideal data structure for our needs.
It is more space-efficient than a sorted vector, since it avoids redundantly storing common prefixes (in a trie, the path to a leaf node implicitly defines the key).
Another advantage of a radix tree is that lookups are in $O(k)$ with $k$ being the key length.
Besides the algorithmic complexity, lookup performance depends on the number of cache misses.
Since we are generally interested in high selectivity and index many cells, the tree usually exceeds cache size (cf., Section~\ref{sec:evaluation}).
Thus, traversing the tree results in many cache misses (at least one per non-cached node).

Let $k_{\mathit{avg}}$ be the average key length and $f$ be the fanout of the tree.
Then the average costs $c_{\mathit{avg}}$ of a lookup can be estimated as follows:
\begin{center}
$c_{\mathit{avg}}$ = $\ceil{k_{\mathit{avg}}/log_2(f)}$ * \textit{costs per node access}
\end{center}
Increasing the number of indexed cells does not necessarily increase $k_{\mathit{avg}}$.
Thus, the costs of a lookup in a non-cached tree are relatively independent of the number of indexed cells.
The number of node accesses is bounded by the maximum key length $k_{\mathit{max}}$,
which is 60 when using S2's adaptive grid index\footnote{The three face bits are indexed in a dedicated \emph{face node} and the level encoding bit is not required.}.
Our tree uses a default fanout $f$ ($f \geq 2$) of 256 (= 8 bits).
Thus every level in the tree represents four S2 levels (recall that every S2 level is encoded with two bits).
This has the side effect that we can only index cells at certain levels.

Let $g$ be the S2 level granularity of the tree.
Then the following holds for indexed cells:
\begin{center}
$\mathit{level}_{\mathit{cell}}$ mod $g$ = 0
\end{center}
Thus, we need to denormalize\footnote{Denormalizing a cell to a given level means replacing the cell with all of its descendant cells at that level.} cells upon insertion and replicate their payloads.

While a fanout of 256 results in sparse nodes and thus in a large space consumption, it allows for efficient lookups.
An adaptive radix tree (ART)~\cite{leis2013adaptive} is usually more space-efficient, however, experiments show that introducing a second node type with four children (Node4 in ART) would (i) only save a negligible amount of space in the case of our workload and (ii) switching between the different node types would have a significant impact on lookup performance.
Also, lookups in compressed node types would be more expensive due to the additional indirection for accessing the payloads.
Since ACT is designed as a transient data structure that is built at runtime, we favor performance over space consumption.
With $r = 256$, the maximum number of node accesses is $\ceil{60/log_2(256)} = 8$.
In practice, a lower $k_{\mathit{max}}$ is often sufficient.
For example, $k_{\mathit{max}} = 48$ allows for indexing S2 cells up to level 24 (a level 24 cell represents at most 2\,$m^2$ on earth)
and limits the number of node accesses to 6.

\begin{figure}[t!]
  \centering

  \begin{subfigure}[b]{0.55\linewidth}
    \includegraphics[width=\linewidth]{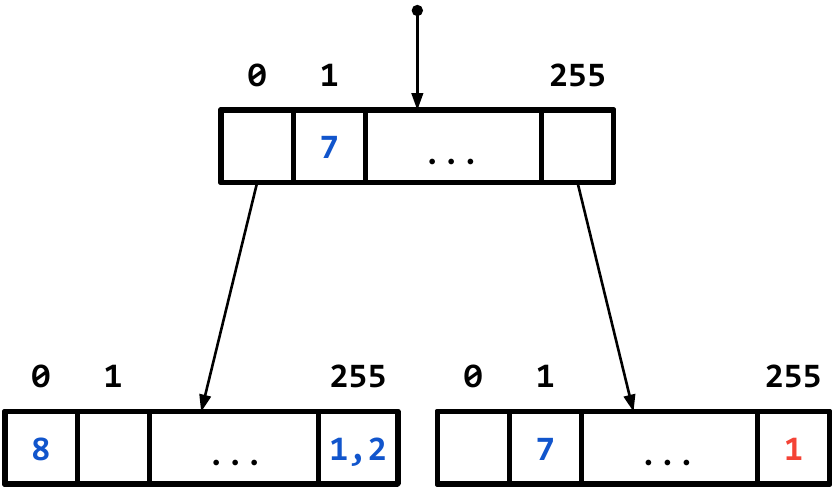}
    \caption{Adaptive Cell Trie}
    \label{fig:radix_tree}
  \end{subfigure}
  ~
  \begin{subfigure}[b]{0.35\linewidth}
    \resizebox{\columnwidth}{!}{
    \begin{tabular}{@{}lll@{}}
    \toprule
    offset & true & candidate \\ \midrule
    0   & \{5\}        & \{3, 1\}         \\
    \mbox{\color{redmaterial} \textbf{1}}   & \{7, 2\}     & \{8\}            \\
    ... & ...          & ...               \\ \bottomrule
    \end{tabular}
    }
    \caption{Lookup table}
    \label{fig:lookup_table}
  \end{subfigure}
  \caption{Adaptive Cell Trie and the lookup table. Key parts (bit sequences)
  are marked in black and values are marked in blue (single or double
  payloads) and red (offsets).}
  \label{fig:radix_lookup}
  \vspace{-1.0em}
\end{figure}

Figure~\ref{fig:radix_tree} illustrates the structure of ACT.
Values (payloads or offsets) can be found in any node at any level of the tree.
Every node consists of a fixed-sized array of 256 entries of 8 byte pointers.
These pointers are \emph{tagged}.
By default, all entries point to a sentinel node indicating a false hit.
Such a tagged entry can be:
\begin{itemize}
  \item An 8 byte pointer to a child or the sentinel node
  \item An inlined payload (a 31 bit value)
  \item Two inlined payloads (two 31 bit values)
  \item An offset (a 31 bit value) into a lookup table
\end{itemize}
We use the two least significant bits of the 8 byte pointer to differentiate between these four possibilities.
For an inlined payload, we differentiate between a true hit and a candidate hit using the least significant bit of the 31 bit payload.
Thus, we can effectively only store $30$ bit payloads (i.e., index up to $2^{30}$ polygons).
We decided to inline two payloads since it is very likely that a cell has less than three payloads in the case of disjoint polygons.
We do not inline more than two payloads since this would mean further reducing the supported number of polygons.

Since our implementation uses S2, we need to maintain up to six radix trees,
one for each face of the cube (cf., Section~\ref{sec:background}).
We call the node that stores the root nodes of these trees \emph{face node}.

{\ParHead Lookup table:}
When a cell references more than two polygons, the tree contains an offset
into a lookup table.
Cells often reference the same set of polygons.
To avoid redundancy, we therefore only store \emph{unique} polygon reference lists.
Figure~\ref{fig:lookup_table} shows an example of a lookup table.
The reference lists are split into two parts, a list with true hits and a list with candidate hits.
Both lists contain polygon ids.
When a cell references at most two polygons, we inline its payloads into the tree.
Assuming disjoint polygons and a uniform distribution of points, a more fine-grained super covering may reduce the chance that we need to access the lookup table.
Let $a$ be the area that is covered by cells that contain an offset into the lookup table (i.e., cells that reference more than two polygons).
Increasing the granularity of the index may reduce the size of $a$ ($\mathit{size}_1 \leq \mathit{size}_0$) and thus reduces the likelihood of having to access the lookup table.
The lookup table is encoded as a single 32 bit unsigned integer array.
The offsets stored in the tree are simply offsets into that array.
Each encoded entry contains the number of true hits followed by the true hits,
the number of candidate hits, and the candidate hits.

\subsection{Training}\label{sec:training}
When we cannot build an index that satisfies a user-defined precision without
exceeding the memory budget, we use an approach that may enter the expensive
refinement phase.
To minimize the likelihood of refinements, we train the index to adapt to the
expected distribution of query points.

The index can either be trained online or offline.
In this work, we only discuss and evaluate how to train the index in a
dedicated training phase.
The online case introduces additional concurrency and buffer management
issues that we leave for future work.

Training the index minimizes the area that is covered by \emph{expensive} cells.
We define expensive cells as cells that map to polygon reference lists
with at least one candidate hit.
When we hit such a cell during the join, we need to enter the refinement phase.
Expensive cells are marked blue and red in Figure~\ref{fig:supercovering}.

We start off with an instance of ACT that does not exceed the memory budget.
When a training point hits an expensive cell, we determine the logical
representation of this cell\footnote{Recall that ACT uses a fanout of 256 thus
the physical cell in the tree might not be the same as the logical cell
that was originally inserted into the tree.}.
Then we verify for each of its four children cells whether they intersect,
are fully contained in, or do not intersect the referenced polygons at all,
and update ACT accordingly.
We repeat this procedure until we have exhausted the memory budget.
We show the effect of training the index in Section~\ref{sec:evaluation}.

\subsection{Probe Index}\label{sec:probe}
A lookup in ACT returns at most one cell that maps to a list of polygon references.
The probe algorithm is shown in Listing~\ref{alg:proberadix}.
In contrast to a search in a binary tree, a search in a radix tree is comparison-free.
This means that we do not compare the value of the search key to the value(s) stored in the current node.
We only need to extract the relevant bits of the search key and jump to the corresponding offset.
However, we do a comparison for checking whether the entry contains a payload.
\noindent\begin{figure}
  \begin{mdframed}[outerlinewidth=0.1pt,outerlinecolor=black,
    innerleftmargin=2pt,innerrightmargin=2pt,innertopmargin=-15pt,innerbottommargin=10pt]
{\small{\small
\begin{lstlisting}[caption={Probe Adaptive Cell Trie},label={alg:proberadix},style=simd,columns=fullflexible]
input:
  face node (*$\codevar{faceN\kern-.1em ode}$*)
  the cell id of the point (*$\codevar{cellId}$*)
output:
  tagged entry (*$\codevar{taggedEntry}$*)
begin
  (*$\codevar{rootNode}$*) = extract face bits of (*$\codevar{cellId}$*) and look up root node in (*$\codevar{faceN\kern-.1em ode}$*)
  if ((*$\codevar{rootN\kern-.1em ode}$*) does not exist)
    // no indexed cells on that face
    return invalid entry
  if (common prefix of (*$\codevar{rootN\kern-.1em ode}$*) does not match)
    return invalid entry
  (*$\codevar{level}$*) = 0
  (*$\codevar{currN\kern-.1em ode}$*) = (*$\codevar{rootN\kern-.1em ode}$*)
  (*$\codevar{bits}$*) = getBits((*$\codevar{cellId}$*), (*$\codevar{level}$*)(*$\textrm{\texttt{++}}$*)) // extract relevant bits
  // traverse the tree until we either hit the sentinel node or found a payload
  while ((*$\codevar{taggedEntry}$*) = (*$\codevar{currN\kern-.1em ode}$*).getEntry((*$\codevar{bits}$*)) is a pointer) {
    if ((*$\codevar{taggedEntry}$*) points to the sentinel node)
      return f(*$ $*)alse hit
    (*$\codevar{currN\kern-.1em ode}$*) = (*$\codevar{taggedEntry}$*)
    (*$\codevar{bits}$*) = getBits((*$\codevar{cellId}$*), (*$\codevar{level}$*)(*$\textrm{\texttt{++}}$*))
  }
end
\end{lstlisting}}}
\end{mdframed}
\vspace{-1.0em}
\end{figure}

For the tagged entry returned by the tree probe, we need to differentiate between (i) one payload,
(ii) two payloads, and (iii) an offset.
In the first case, we need to check whether the payload is invalid, which indicates a false hit.
Otherwise, we extract the interior flag (the least significant bit of the 31 bit payload) and the polygon id and return a polygon reference.
In the second case, we extract and return both references.
Only in the third case, we need to access the lookup table to retrieve the
polygon references.

Listing~\ref{alg:probe} shows the complete probe algorithm.
For a given point, we retrieve the cell that it is contained in (if such a cell exists)
and go over all references of this cell.
We can skip refinement checks for true hits and immediately output the join partners.

\noindent
\begin{minipage}{\linewidth}
  \vspace{0.5em}
  \begin{mdframed}[outerlinewidth=0.1pt,outerlinecolor=black,
    innerleftmargin=2pt,innerrightmargin=2pt,innertopmargin=-15pt,innerbottommargin=10pt]
{\small{\small
\begin{lstlisting}[caption={The probe algorithm},label={alg:probe},style=simd,columns=fullflexible]
input:
  points (*$\codevar{points}$*) // lat/lng coordinates and cell ids
  polygons (*$\codevar{polygons}$*) // lat/lng coordinates of vertices
  face node (*$\codevar{faceN\kern-.1em ode}$*)
  lookup table (*$\codevar{lookup\kern-.1em Table}$*)
output:
  list of join pairs (*$\codevar{pairs}$*) // point/polygon pairs
begin
  for ((*$\codevar{point}$*) in (*$\codevar{points}$*)) {
    (*$\codevar{tagged\kern-.1em Entry}$*) = probeAdaptiveCellTrie((*$\codevar{face\kern-.1em Node}$*), (*$\codevar{point.cellId}$*)) // (*$\mbox{\color{mygreen} cf., Listing \ref{alg:proberadix}}$*)
    if ((*$\codevar{taggedEntry}$*) is invalid)
      continue
    (*$\codevar{references}$*) = getPolygonReferences((*$\codevar{lookup\kern-.1em Table}$*), (*$\codevar{taggedEntry}$*)) // (*$\mbox{\color{mygreen} returns a list of polygon references}$*)
    for ((*$\codevar{reference}$*) in (*$\codevar{references}$*)) {
      (*$\codevar{polygonId}$*) = (*$\codevar{reference.polygonId}$*)
      (*$\codevar{polygon}$*) = (*$\codevar{polygons[polygonId]}$*)
      if ((*$\codevar{reference}$*) is t(*$ $*)rue hit) {
        add {(*$\codevar{point}$*), (*$\codevar{polygon}$*)} to (*$\codevar{pairs}$*)
      } else { // candidate hit
        if (polygonCoversPoint((*$\codevar{polygon}$*), (*$\codevar{point}$*))) // exact PIP test
          add {(*$\codevar{point}$*), (*$\codevar{polygon}$*)} to (*$\codevar{pairs}$*)
  } } }
end
\end{lstlisting}}}
\end{mdframed}
\end{minipage}
\vspace{1em}

\section{Modern Hardware}\label{sec:simd}
In this section, we present the optimizations of our proposed join algorithm
for modern hardware architectures.
The general hardware trend is having an increasing number of cores with ever
larger vector processing units (VPUs).
That applies to established server-class machines such as Intel's Xeon
and particularly to high-performance computing platforms such as Intel's Xeon Phi.
In this work, we focus on Intel's second-generation Xeon Phi processor,
Knights Landing.

\subsection{Knights Landing Processor}\label{sec:knl}
Just like its predecessor, the KNL processor is a \emph{many
integrated core} (MIC) architecture which draws its computational power from
wide VPUs.
The corresponding AVX-512 offer SIMD capabilities on 512-bit registers.
In contrast to the first generation Phi, KNL is available as
a co-processor expansion card \emph{and} as a self-boot processor (socket on a
mainboard) which is fully backward compatible as it also implements all legacy
instructions.
In this work, we focus on the self-boot processor and \emph{not} the co-processor
expansion card.

Compared to our Xeon processor (cf., Section~\ref{sec:evaluation}),
equipped with 14 fast clocked \emph{brawny} cores per socket,
the KNL processor consists of 64
\emph{wimpy} cores moderately clocked at 1.3\,GHz.
The register width as well as the number of vector registers
has been doubled over AVX2 architectures.
Thus, for a software to run efficiently on KNL, it is essential to exploit the
offered SIMD capabilities.

Especially for data-intensive workloads like ours, KNL offers three promising
features:
(i) it is equipped with an on-chip High Bandwidth Memory (HBM) which is
comparable to the memory on high-end GPUs in speed and size,
(ii) the processor performs a much more aggressive memory prefetching,
and (iii) the 4-way hyper threading can help to hide memory latencies.

\subsection{Parallelism}
KNL is a massively parallel processor in terms of thread-level parallelism
and data parallelism on the instruction level.
\subsubsection{Multithreading}
Thread-level parallelism can be achieved using well-known techniques.
In existing parallel joins, multiple threads probe an index
structure (e.g., a hash table) in parallel.
During the probe phase, the index structure is in an immutable state.
Therefore, an arbitrary number of threads can perform read operations without
the need for synchronization.
As no contention points exist, the join performance typically scales linearly
in the number of threads.

\subsubsection{Data Parallelism}
Achieving data parallelism on instruction level is more involved and
requires several modifications over the scalar implementation.
The goal is again to perform multiple index lookups in parallel.
We divide a 512 bit SIMD register into eight 64 bit SIMD lanes and
can thus process up to eight elements simultaneously per vector unit.
The major difference to thread parallelism is, that the same instruction is
executed on all in-flight lookups, which is due to the lockstep nature of SIMD.
Thus, for the SIMD implementation to be efficient, the in-flight lookups need
to make progress with every issued instruction.
This implies that every lookup must be in the same \emph{stage} of the algorithm,
i.e., it is no longer possible that one lookup performs a prefix comparison
while simultaneously another issues a memory load.
To achieve this, our probe algorithm from Listing~\ref{alg:proberadix} is decomposed
into the following stages:

{\ParHead Determine tree root:}
As mentioned earlier, our implementation uses S2 for space partitioning
the earth.
Thus, a join index actually consists of up to six radix trees.
In this stage, we extract the three face bits of the query point's cell id
to determine the radix tree that needs to be traversed.
Further, we check if the query point and the polygons indexed in the radix tree
share the predetermined \emph{common prefix}.
If the prefixes do not match, then it is guaranteed that there is no join partner.
We want to point out that on the one hand this stage of the algorithm is S2
specific but on the other hand it is similar to a scenario where the data
is radix partitioned by three bits.

{\ParHead Tree traversal:}
In this stage, we traverse the tree.
In each iteration, we compute the offset of the entry in the current tree node
according to the search key.
If the entry is a pointer to a child tree node we follow the pointer
and continue.

{\ParHead Produce output:}
In the third and last stage, we interpret the content of the entries where
the search in the previous stage terminated.
We have to distinguish the cases (i) false hit, (ii) single or double hit,
and (iii) multiple hits which are either true or candidate hits (cf., Section~\ref{sec:approach}).

\vspace{1em}

To illustrate the vectorized implementation, we use pseudo code instead of C\texttt{++}
for better readability.
In the pseudo code, vectors are annotated with an arrow, bitmasks are named
|m|$_{\mbox{\lstinline[basicstyle=\scriptsize\sffamily]{subscript}}}$, and for
\emph{masked operations} we use the following notation:
\begin{center}
$\vv{\mbox{\lstinline[style=simd]{dst}}}$ =
\,|operation|$_{\mbox{\lstinline[basicstyle=\scriptsize\sffamily]{mask}}}^{\vv{\hbox{\lstinline[basicstyle=\scriptsize\sffamily]{src}}}}$|(|$\vec{\hbox{\small\lstinline[style=simd]{a}}}$|)|
\end{center}
The |operation| is applied to the components in
$\vec{\mbox{\lstinline[style=simd]{a}}}$ specified by \lstinline[style=simd]{mask}
and the results are stored in $\vv{\mbox{\lstinline[style=simd]{dst}}}$.
The remaining components are copied from
$\vv{\mbox{\lstinline[style=simd]{src}}}$ to $\vv{\mbox{\lstinline[style=simd]{dst}}}$.
Further, we make use of the |vpermd/q| instruction (for better readability we refer to it as |permute|)
that shuffles vector components using the corresponding index vector, e.g.,
\begin{center}
$\underbrace{[\text{d,a,d,b}]}_{\textrm{result vector}}$ = |permute(|$\underbrace{[\text{3,0,3,1}]}_{\textrm{index vector}}, \underbrace{[\text{a,b,c,d}]}_{\textrm{input vector}}$|)|.
\end{center}

\noindent
During index construction we initialize the following vectors:
\begin{mdframed}[outerlinewidth=0.1pt,outerlinecolor=black,
  innerleftmargin=2pt,innerrightmargin=2pt,innertopmargin=3pt,innerbottommargin=10pt]
{\small{\small
\begin{lstlisting}[style=simd,columns=fullflexible]
  (*$\codevec{roots} = [ \text{r}_0, ... , \text{r}_5 , \_ , \_  ]$*) // pointers to the six root nodes
  (*$\codevec{prefixes} = [ \text{p}_0, ... , \text{p}_5 , \_ , \_ ]$*) // common prefixes
  (*$\codevec{prefix\_lengths} = [ \text{l}_0, ... , \text{l}_5 , \_ , \_ ]$*) // common prefix lengths
\end{lstlisting}
}}
\end{mdframed}
\vspace{-0.5em}

\noindent
These vectors are used to determine the tree roots in the first stage
(cf., Listing~\ref{algvec:stage1}).
-- For performance reasons these vectors remain in CPU registers to avoid memory loads. --
Given a vector of cell ids, we first extract the face bits of each cell id.
The resulting vector is then used as a permutation index vector to
move the pointer of the corresponding root nodes to the output vector $\codevec{\small\textsf{nodes}}$.
Similarly, we obtain the prefixes and the lengths of the prefixes.
The actual prefix check, an equality comparison, results in a bitmask
which is used as an execution mask in the later stage.
I.e., if the cell id in SIMD lane $i$ does not match the common prefix, then the
$i$-th bit in the bitmask is set to zero.
\noindent\begin{figure}
  \begin{mdframed}[outerlinewidth=0.1pt,outerlinecolor=black,
    innerleftmargin=2pt,innerrightmargin=2pt,innertopmargin=-15pt,innerbottommargin=10pt]
\begin{lstlisting}[style=simd,columns=fullflexible,label={algvec:stage1},caption={Stage 1 of the vectorized probe algorithm}]
input:
  (*$\textrm{cell id vector}$*) (*$\codevec{cell\_ids}$*)
output:
  (*$\textrm{tree node pointers}$*) (*$\codevec{nodes}$*)
  (*$\textrm{active lane mask}$*) m
begin
  (*$\codevec{faces}$*) = (*$\codevec{cell\_ids} \gg 61$*) // extract the 3 face bits
  (*$\codevec{nodes}$*) = permute((*$\codevec{faces}$*), (*$\codevec{roots}$*)) // get the node pointers
  // get common prefix and prefix length for each tree
  (*$\codevec{p}$*) = permute((*$\codevec{faces}$*), (*$\codevec{prefixes}$*))
  (*$\codevec{l}$*) = permute((*$\codevec{faces}$*), (*$\codevec{prefix\_lengths}$*))
  (*$\codevec{p}_{\text{actual}}$*) = (*$\codevec{cell\_ids} \gg (61 - \codevec{l})$*)  // extract the common prefixes
  m = (*$(\codevec{p}_{\text{actual}}$*) == (*$\codevec{p})$*) // prefix check (results in a bitmask)
end
\end{lstlisting}
\end{mdframed}
\vspace{-1.5em}
\end{figure}

The tree traversal, shown in Listing~\ref{algvec:stage2},
is more involved.
As it performs multiple traversals in lockstep, it is required to
do some bookkeeping about the current traversal state.
The mask |m|$_{\text{\textsf{traverse}}}$ keeps track of the active SIMD lanes
and in |m|$_{\text{\textsf{output}}}$ the algorithm memorizes the lanes that
produced an output.
The output mask therefore represents the execution mask for the next stage.
The output values itself are stored in $\codevec{\small\textsf{values}}$.

In each iteration of the while loop, the algorithm first computes the entry
offsets from which it subsequently \emph{gathers} the contents.
The gather is a masked operation for two reasons:
(i) to avoid unnecessary memory accesses
and (ii) to keep already existing output values in $\codevec{\small\textsf{values}}$.

Once the bucket contents are loaded into a vector register,
we have to check whether the values are payloads or pointers.
In case of pointers, the addresses are compared
to the address of the sentinel node.
If the comparison evaluates to true, the search terminates without producing
an output, otherwise the corresponding pointers are moved to $\codevec{\small\textsf{nodes}}$.
Based on the two masks |m|$_{\text{\textsf{ptr}}}$ and |m|$_{\text{\textsf{sentinel}}}$,
the traverse mask |m|$_{\text{\textsf{traverse}}}$ is updated and the search
continues until |m|$_{\text{\textsf{traverse}}}$ is zero.
\noindent\begin{figure}
  \begin{mdframed}[outerlinewidth=0.1pt,outerlinecolor=black,
    innerleftmargin=2pt,innerrightmargin=2pt,innertopmargin=-15pt,innerbottommargin=10pt]
\begin{lstlisting}[style=simd,columns=fullflexible,label={algvec:stage2},caption={Stage 2 of the vectorized probe algorithm}]
input:
  (*$\textrm{cell id vector}$*) (*$\codevec{cell\_ids}$*)
  (*$\textrm{node pointer vector}$*) (*$\codevec{nodes}$*)
  (*$\textrm{mask}$*) m
output:
  (*$\textrm{output mask}$*) (*$\text{m}_{\text{output}}$*)
  (*$\textrm{output payloads}$*) (*$\codevec{values}$*)
begin
  level = 1 // start tree traversal at level 1
  m(*$_{\text{traverse}}$*) =  m // active lanes
  m(*$_{\text{output}}$*) = 0 // lanes that produced an output
  (*$\codevec{values}$*) = [0, ... , 0] // a place for the produced output
  while (m(*$_{\text{traverse}}$*) != 0) {
     (*$\codevec{offsets} = $*) compute_bucket_offset((*$\codevec{cell\_ids},$*) level)
     // load bucket contents
     (*$\codevec{values} = $*) gather(*$_{\text{m}_{\text{traverse}}}^{\codevec{values}}$*)((*$\codevec{nodes} ~ \texttt{+} ~ \codevec{offsets}$*))
     // check for pointer (the two LSBs are zero)
     m(*$_{\text{ptr}}$*) = (((*$\codevec{values}$*) & 11(*$_\text{2}) ==_{\text{m}_{\text{traverse}}}$*) 0)
     m(*$_{\text{sentinel}}$*) = ((*$\codevec{values} ==_{\text{m}_{\text{ptr}}} $*) sentinel) // check for sentinel
     (*$\codevec{nodes} =_{\text{m}_{\text{ptr}}} \codevec{values}$*) // update current node pointers
     // identify lanes that produced an output
     m(*$_{\text{output}} \,\vert=\,\, \text{!m}_{\text{ptr}}$*) & (*$\text{m}_{\text{traverse}}$*)
     // identify trapped lanes and update traversal mask
     (*$\text{m}_{\text{done}} \,\vert=\,\, \text{m}_{\text{output}}$*) | (*$\text{m}_{\text{sentinel}}$*)
     (*$\text{m}_{\text{traverse}}$*) ^= (*$\text{m}_{\text{done}}$*) & (*$\text{m}_{\text{traverse}}$*)
     level(*$\texttt{++}$*)
  }
end
\end{lstlisting}
\end{mdframed}
\vspace{-1.5em}
\end{figure}

In the last stage, we have to distinguish between true and candidate hits.
We have not vectorized this stage since its performance impact is neglectable.
However, for the |select polygon_id, count(*) ... group by polygon_id| query we
use in the evaluation (cf., Section~\ref{sec:evaluation}), we partially optimized the
aggregation using SIMD. Otherwise, the aggregation would become a bottleneck
due to the wimpy CPU cores of KNL.

Our SIMDfied aggregation is optimized for the most likely cases,
where either one polygon id or two polygon ids are returned by the ACT lookup.
We have implemented a direct aggregation based on \emph{gather} and \emph{scatter}
instructions and the efficient \emph{popcount} implementation presented in~\cite{q1gubner}.
In the less likely cases, where the query point matches more than two polygons
(i.e., ACT returns an offset into the lookup table) or where the match is
inconclusive, we materialize the results into a small L1 resident buffer,
which is consumed sequentially later on.

\section{Evaluation}\label{sec:evaluation}
In this section, we show the performance of our join and the selectivity of
our index for a join between points and polygons on two different machines.
We compare the performance of our technique to an R-tree based approach that
uses the same code in the refinement phase as ours.
To give a point of reference, we also compare against PostGIS and Magellan.
In addition to the results of the exact algorithm, we show that an approximate
algorithm can significantly speed up the join while providing sufficiently
precise results for many use cases.

{\ParHead Configuration:}
We evaluate our approach on a server-class machine (Xeon)
and on KNL (cf., Section~\ref{sec:knl}).
Xeon runs Ubuntu 16.04 and is equipped with an
Intel Xeon E5-2680 v4 CPU (2.40\,GHz, 3.30\,GHz turbo)
and 256\,GB DDR4 RAM.
The machine has two NUMA sockets with 14 physical cores each,
resulting in a total of 28 physical cores (56 hyperthreads).
KNL runs CentOS Linux release 7.2.1511 and is equipped with a single
Intel Xeon Phi CPU 7210 (1.30\,GHz, 1.50\,GHz turbo),
96\,GB DDR4 RAM, and 16\,GB on-chip HBM.
We use GCC version 5.4.0 with \texttt{O3} enabled in all experiments.

We join 1.23\,B points from the NYC taxi dataset\footnote{\url{http://www.nyc.gov/html/tlc/html/about/trip_record_data.shtml}}
against NYC's boroughs (5 polygons), neighborhoods (289 polygons), and census
blocks (39,184 polygons) and count the number of points per polygon.
While there are only five boroughs, their polygons are significantly more
complex.

We compare our technique to an approach that builds an R-tree on the polygons,
probes the points against the index, and refines candidate hits.
In the refinement phase, we use S2's PIP test, the same algorithm that our
technique uses.
In the filter phase, we use the boost R-tree implementation (1.6.0) and evaluate
different splitting strategies (|linear|, |quadratic|, and |rstar|) and maximum
element counts per node (4, 8, 16, and 32).
|rstar| with a maximum of 8 elements per node performs best in all workloads.
We therefore omit the results for the other configurations.
We have also evaluated the |STRtree| implementation in GEOS (3.5.0),
however, omit it in the further evaluation as it cannot compete with |rstar|
for our workloads.
For example in the neighborhoods join on Xeon, |STRtree|
achieves a throughput of 27.4\,M points/s, while |rstar| processes 70.9\,M points/s.
The performance difference is caused by the selectivity of the respective
index structure.
|rstar| returns 1.60 candidates on average while |STRtree| returns 1.94.
In addition, we benchmark PostgreSQL 9.6.1 (PostGIS 2.3.1) with a GiST index
on the polygons.
We configured PostgreSQL to use as many parallel workers\footnote{The intra-query parallelism feature was introduced in PostgreSQL 9.6.}
as available hyperthreads.
As another competitor, we evaluate a single-machine deployment of Magellan.
Similar to our approach, Magellan uses true hit filtering.
Our default configuration for computing the individual coverings is as follows:
max covering cells = 128, max covering level = 30, max interior cells = 256,
and max interior level = 20.

\begin{figure*}[t!]
  \centering

  \begin{subfigure}[b]{0.48\linewidth}
    \includegraphics[width=\linewidth]{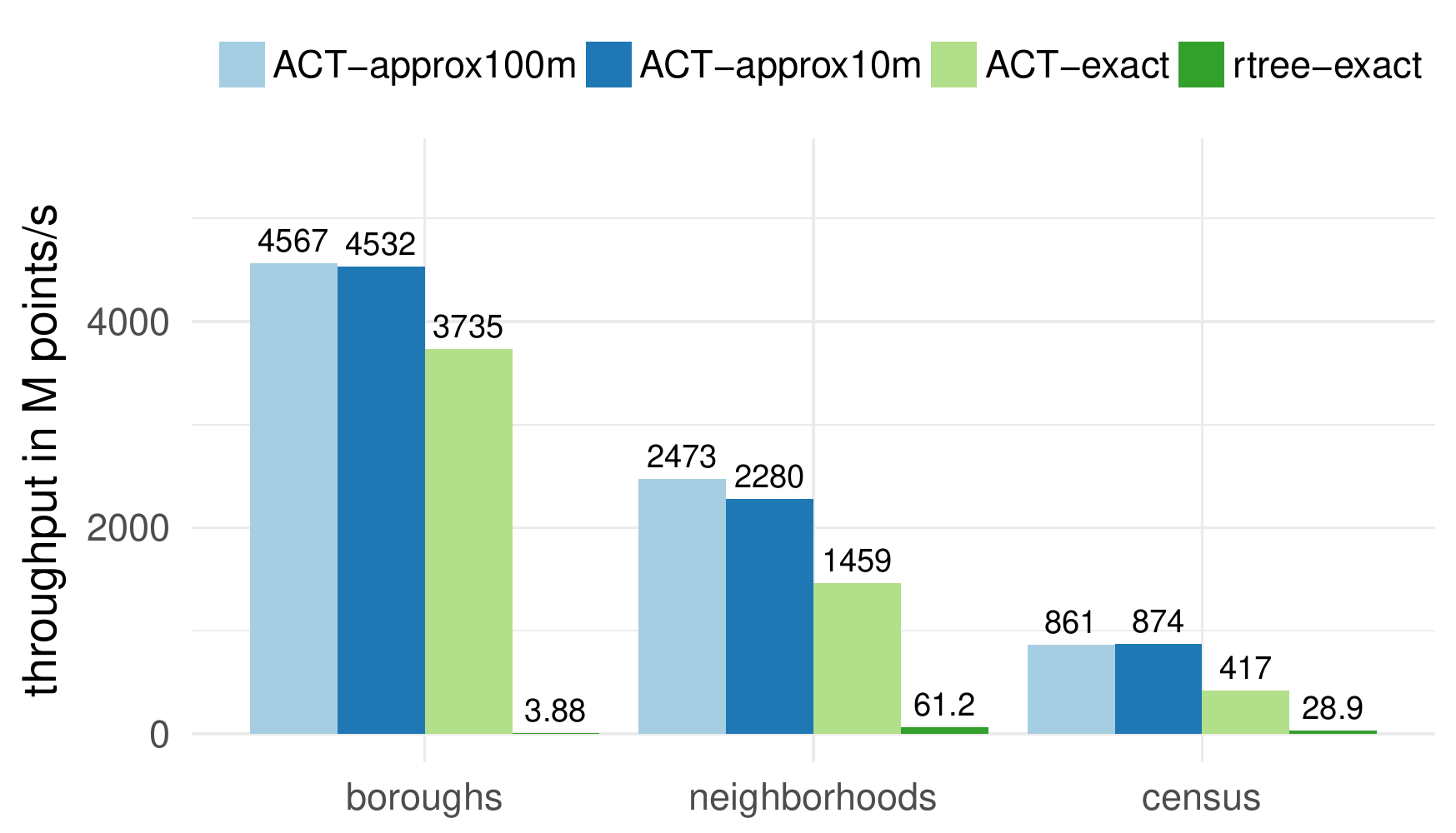}
    \caption{\textbf{Dual-socket} Xeon}
    \label{fig:competitors_xeon}
  \end{subfigure}
  ~
  \begin{subfigure}[b]{0.48\linewidth}
    \includegraphics[width=\linewidth]{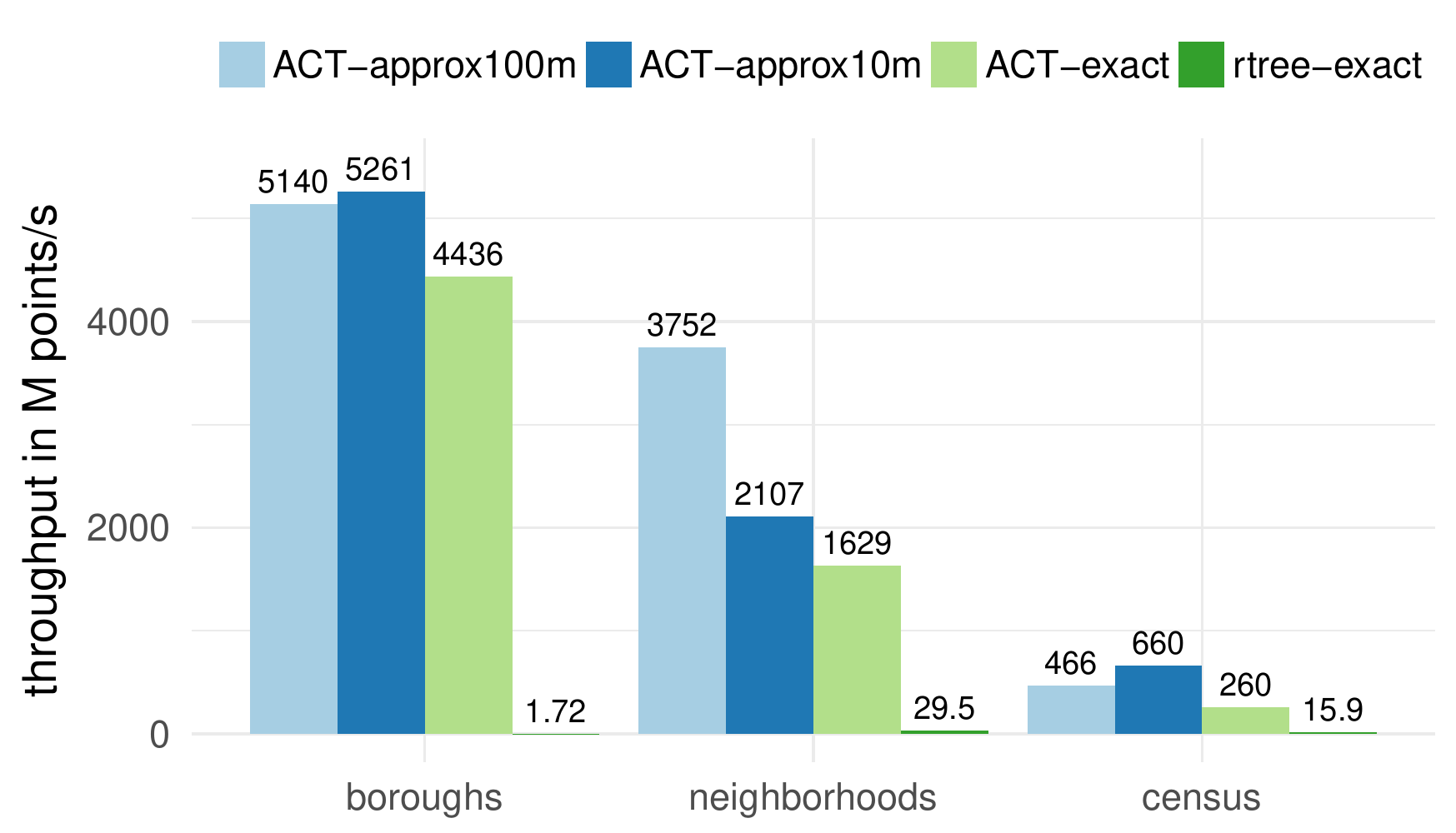}
    \caption{\textbf{Single-socket} KNL}
    \label{fig:competitors_knl}
  \end{subfigure}
  \caption{Throughput of our approach (ACT) compared to the best competitor (boost R-tree) on the Xeon and on the KNL machine
  (\texttt{approx100m} and \texttt{approx10m} = approximate with 100 and 10\,m precision,
  \texttt{exact} = refine candidate hits)}
  \label{fig:competitors}
  \vspace{-1.0em}
\end{figure*}

{\ParHead Results:}
Figure~\ref{fig:competitors} shows the throughput of our approach compared
to the boost R-tree using all threads (56 and 256 threads on Xeon and KNL, respectively).

\texttt{approx} always skips the refinement phase and treats all candidate hits
as true hits.
The precision of \texttt{approx} is bounded by the diagonal length of the largest
covering cell.
We refined the index to only contain sufficiently small covering cells to
guarantee the denoted precision.
\texttt{approx10m} has a false positive rate of almost 0\% for boroughs,
3\% for neighborhoods, and 17\% for the census blocks dataset.
The average distance of these false positives join partners is 1.31\,m for boroughs,
1.62\,m for neighborhoods, and 1.50\,m for the census blocks dataset.
This shows that our technique can achieve a significantly better precision
than the user-defined precision bound (10\,m in that case) for real workloads.
\texttt{approx10m} consumes 1.2\,GB of memory for the largest of the three
datasets.

\texttt{exact} denotes the full join that refines candidate hits using S2.
We train the index with 1\,M historical data points for all datasets.

\begin{figure}
\centering
\includegraphics[width=1\linewidth]{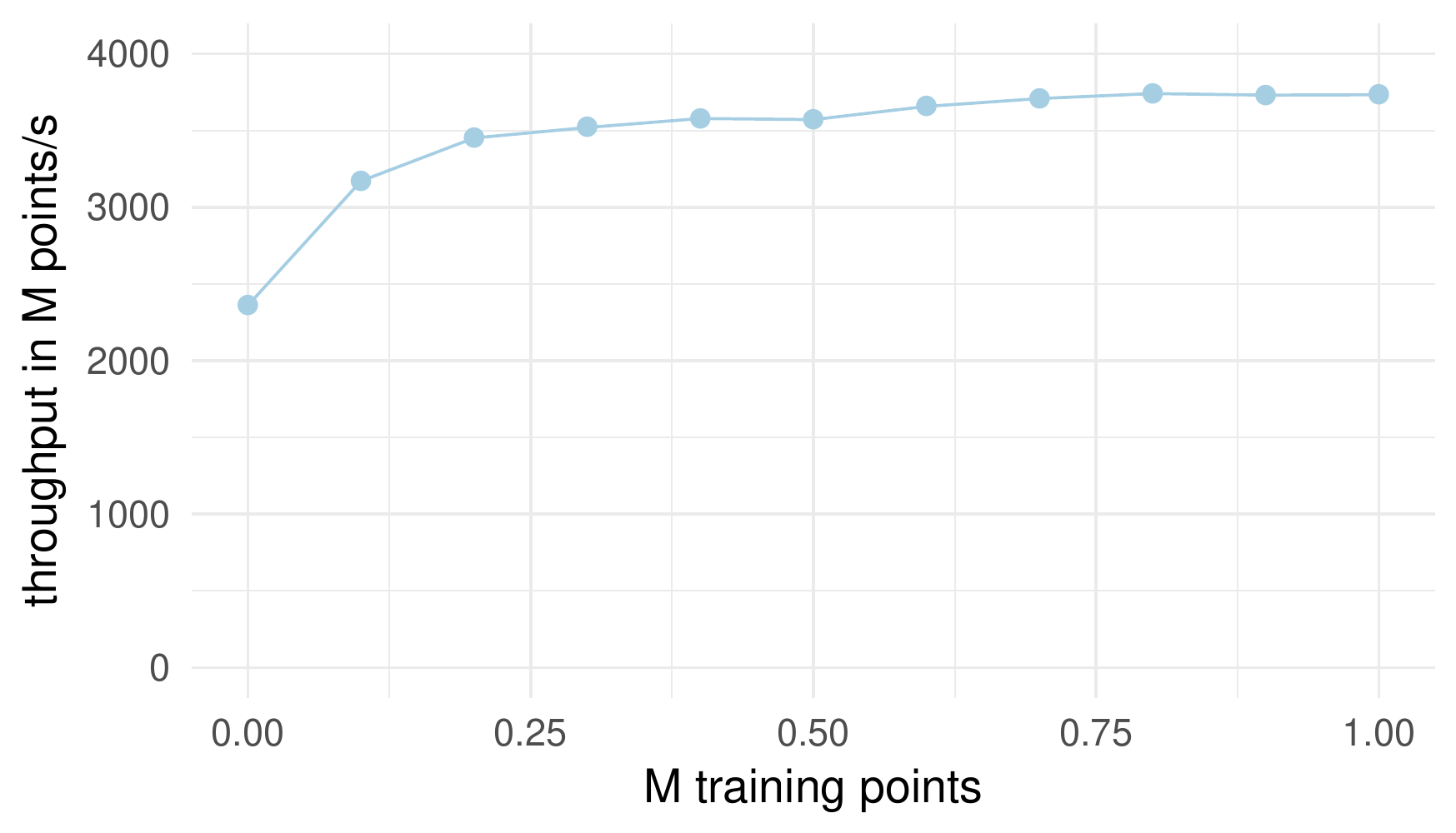}
\caption{Throughput after training ACT on boroughs with an increasing
number of historical data points}
\label{fig:training}
\vspace{-1.0em}
\end{figure}

Figure~\ref{fig:training} shows the throughput increase when training ACT with
an increasing number of points.
An untrained ACT achieves a throughput of 2360\,M points/s, whereas ACT
trained with 1\,M points achieves 3732\,M points/s (+58\%).

The Xeon binaries are compiled with AVX2 optimizations (\texttt{march=core-avx2})
while the KNL implementation is hand-optimized (cf., Section~\ref{sec:simd}).
Compared to an auto-vectorized version, the performance of our hand-optimized
implementation is between 29\% and 70\% higher.
Auto-vectorization with AVX-512 (\texttt{march=knl}) instead of AVX2 does
not have a performance impact suggesting that hand-tuning is essential to fully
utilize the wide vector processing units of KNL.

On Xeon (Figure~\ref{fig:competitors_xeon}), \texttt{ACT-exact} achieves a throughput
of 3735\,M points/s for the boroughs dataset.
With a higher number of polygons in the neighborhoods and census datasets,
the throughput of \texttt{ACT-exact} decreases.
The reason for the slowdown is the lower selectivity and the higher space consumption
of our index for the larger datasets (cf., Table~\ref{tbl:metrics}).
While the index of the boroughs dataset fits into the cache, the index of the
neighborhoods and census blocks datasets exceed the cache size.
The R-tree achieves a peak performance of 61.2\,M points/s for the
neighborhoods dataset.
The reason for its slow performance for the boroughs dataset is as follows:
S2's PIP test implements the ray tracing algorithm.
As mentioned earlier, this algorithm has a runtime complexity of $O(n)$ with
$n$ being the number of edges.
Since the boroughs are complex polygons with many edges, the PIP tests in the
refinement phase are very expensive.
Here, our algorithm shines since it can identify most join partners in the
filter phase and only needs to enter the refinement phase for 0.1\% of the points.

As a point of reference, PostgreSQL (PostGIS) with a GiST index on the polygons
only achieves a throughput of 0.39\,M points/s for the boroughs,
1.09\,M points/s for the neighborhoods, and
0.69\,M points/s for the census blocks dataset.
Similar to the R-tree based join, PostgreSQL's performance suffers from the
complex polygons in the boroughs dataset.
Even though PostgreSQL parallelizes this query, its performance is still
orders of magnitude lowers than ours.
Magellan performs better than PostGIS and achieves a throughput of
0.88\,M points/s for boroughs, 4.57\,M points/s for neighborhoods, and
2.24\,M points/s for census blocks.

With 10 meter precision, ACT achieves a throughput of 4532\,M points/s for the
boroughs dataset and still 874\,M points/s for the almost 8000x larger census
blocks dataset.
The reason for the similar performance of \texttt{ACT-approx100m} and
\texttt{ACT-approx10m} is that neither of the two performs any exact checks.
Thus, their performance is dominated by the costs for the ACT node accesses
and the final aggregation.
For the census blocks dataset, \texttt{approx10m} is even faster than
\texttt{approx100m}, since the finer-grained index of \texttt{approx10m}
returns less candidates on average leading to less aggregations.

On KNL (Figure~\ref{fig:competitors_knl}), \texttt{ACT-exact} achieves a peak
throughput of 4436\,M points/s for the boroughs dataset outperforming the
Xeon counterpart.
For the census blocks dataset, we can see the same effect as on the Xeon machine
with the more precise \texttt{approx} approach achieving the highest throughput.
These results show that the single-socket KNL can compete with or even outperform
a dual-socket Xeon machine when the code uses vector instructions.

\begin{table}
\centering
\caption{Metrics of our index}
\label{tbl:metrics}
\begin{tabular}{@{}llll@{}}
\toprule
metric                         & boroughs        & neighborhoods   & census          \\ \midrule
tree nodes                     & 413             & 13025           & 590147          \\
false hits                 & 0.07\%          & 0.09\%          & 0.07\%          \\
\textbf{solely true hits} & \textbf{99.9\%} & \textbf{87.1\%} & \textbf{72.1\%} \\
candidates                     & 1.06            & 1.66            & 1.72            \\ \bottomrule
\end{tabular}
\vspace{-1.0em}
\end{table}

\begin{table}
\centering
\caption{Effect of training the index}
\label{tbl:metrics_adaptive}
\begin{tabular}{@{}llll@{}}
\toprule
metric                         & boroughs        & neighborhoods   & census          \\ \midrule
tree nodes                     & 456             & 22484           & 634005          \\
false hits                 & 0.07\%          & 0.11\%          & 0.07\%          \\
\textbf{solely true hits} & \textbf{99.9\%} & \textbf{97.7\%} & \textbf{88.6\%} \\
candidates                     & 1.15            & 1.57            & 1.70            \\ \bottomrule
\end{tabular}
\vspace{-1.0em}
\end{table}

\begin{figure*}[t!]
  \centering

  \begin{subfigure}[b]{0.48\linewidth}
    \includegraphics[width=\linewidth]{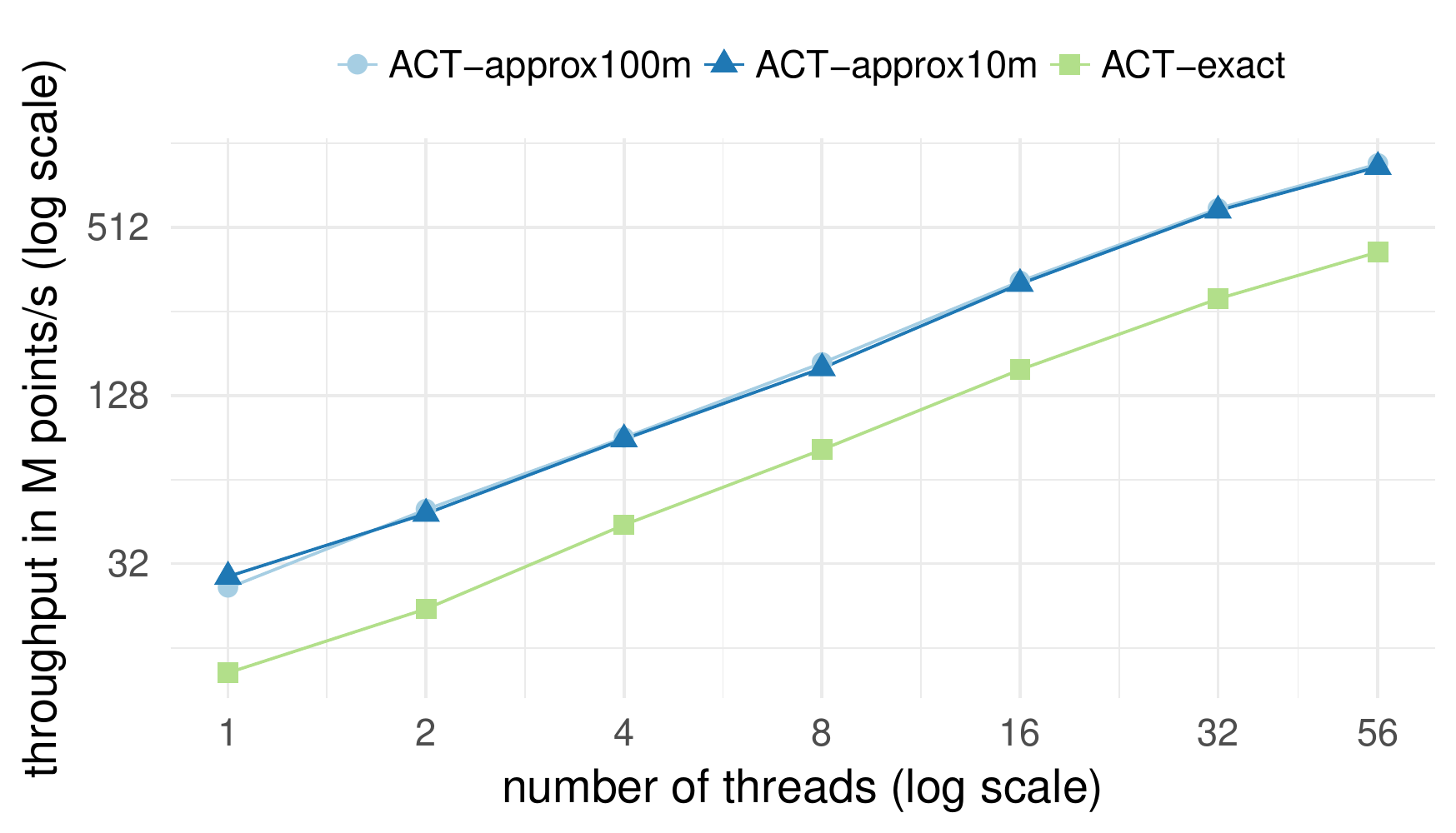}
    \caption{Dual-socket Xeon}
    \label{fig:scalability_xeon}
  \end{subfigure}
  ~
  \begin{subfigure}[b]{0.48\linewidth}
    \includegraphics[width=\linewidth]{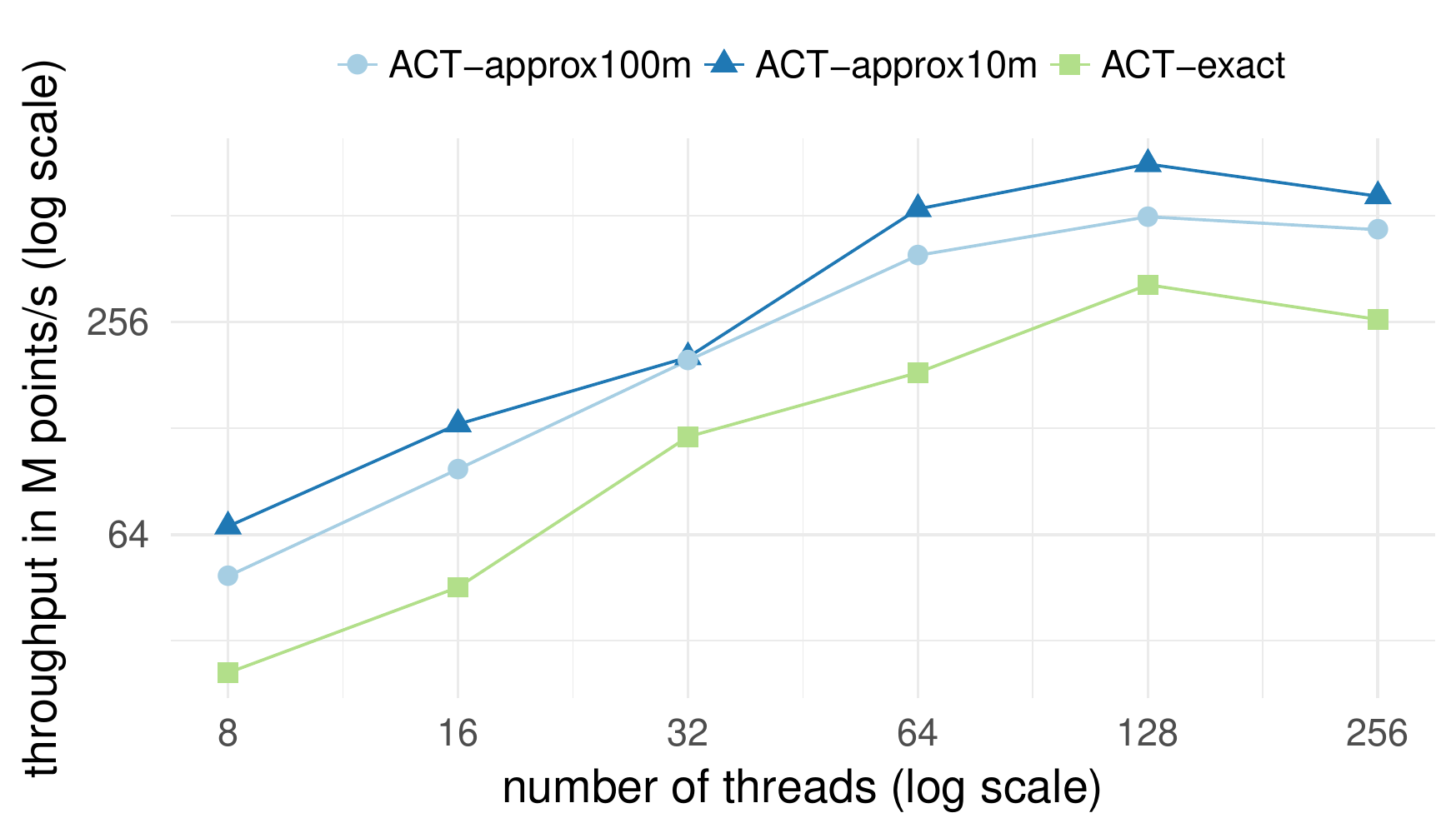}
    \caption{Single-socket KNL}
    \label{fig:scalability_knl}
  \end{subfigure}
  \caption{Scalability for the census blocks dataset
  on the Xeon and on the KNL machine}
  \label{fig:scalability}
  \vspace{-1.0em}
\end{figure*}

Table~\ref{tbl:metrics} shows different metrics of our index for the three
polygon datasets.
The quality metric \emph{solely true hits} indicates the percentage of points
that skipped the expensive refinement phase,
which is clearly above 70\% throughout all datasets.
The points for which the index probe neither returns a false hit nor solely
true hits and thus need to enter the refinement phase are compared against less
than two candidate polygons on average.
For the boroughs dataset, our index consumes less than one MB of
memory and thus the top levels of the tree fit into the L2 cache of the Xeon machine.
For the larger datasets, we use a larger number of cells and thus also a larger
index structure.
The \texttt{exact} index of the census blocks dataset requires more than one GB
of memory.
The high memory consumption is caused by the high fanout of 256 of our radix tree.
An adaptive radix tree would be more space-efficient, however, switching
between the different node types has a significant impact on probe performance,
especially for cache-resident indices.

Table~\ref{tbl:metrics_adaptive} shows the effect of training the index
with 1\,M training points (cf., Section~\ref{sec:logicalindex}).
This optimization improved the percentage of solely true hits for all three
datasets (especially for census blocks) while increasing the number of tree
nodes and thus the space consumption of the indices.

Finally, we evaluate the scalability of our technique on both machines.
We choose the census blocks dataset for this experiment since its index
significantly exceeds the cache sizes of both machines but still fits into the MCDRAM of KNL.
Figure~\ref{fig:scalability} shows the results.

On Xeon, all three approaches scale almost linearly with the number of physical cores
and benefit from hyperthreading.
The fact that an oversubscription of cores has a positive performance impact
shows that our technique is bound by memory access latencies and having more
threads than physical cores can hide these latencies.

On KNL, \texttt{ACT-approx10} has its maximum performance with 2-way hyperthreading
(128 threads) and decreases with 4-way hyperthreading (256 threads).
\texttt{ACT-exact} scales from 13.0\,M points/s with one thread to 417\,M
points/s with 56 threads on Xeon.
On KNL, \texttt{ACT-exact} scales from 25.9\,M points/s with eight threads to
260\,M points/s with 256 threads.

\section{Related Work}\label{sec:relatedwork}
Geospatial joins have been studied for decades and there is a lot of related
work on algorithmic techniques.
\cite{jacox2007spatial} gives a summary of spatial join techniques.
\cite{brinkhoff1994multi} proposes true hit filtering in the form of maximum
enclosed rectangles and circles allowing to skip the refinement phase in many cases.
\cite{zimbrao1998raster} follows up on this approach by using raster approximations
in the form of uniform grids thereby improving the selectivity.
Both approaches optimize not only for selectivity but also for I/O operations,
which is an important performance factor for disk-based database systems.
\cite{kothuri2001efficient} recursively divides the MBR of a polygon into four
cells until a certain granularity is reached, identifies interior cells, and
indexes them in an R-tree to skip refinement checks during join processing.
\cite{roth09:esi} presents an approach that leverages multiple grids
with different resolutions to index geospatial objects using standard database indices.

In contrast to existing techniques, our approach proposes a new way of
processing geospatial joins by identifying the majority or even all of the join partners
in the filter phase using an adaptive grid \emph{while} leveraging the large main-memory capacities,
high memory bandwidths, multi-cores, and wide vector processing units of modern hardware.
None of these techniques supports training the index with historical data
points to improve probe performance.

Several database systems and geographical information systems (GIS) support geospatial joins.
\cite{eldawy2015era} present a comprehensive survey of state-of-the-art systems
and classifies the approaches that exist in literature to design and implement such systems.
There are two main geospatial processing libraries used by database systems: GEOS\footnote{\url{http://trac.osgeo.org/geos/}} and S2 (cf., Section~\ref{sec:background}).
MongoDB, a document-oriented database, uses S2 for geospatial processing and indexes the coverings of geospatial objects in a B-tree\footnote{\url{http://blog.mongodb.org/post/50984169045/new-geo-features-in-mongodb-24}}.
HyPerSpace~\cite{pandey2016high}, a geospatial extension to the main-memory database system HyPer~\cite{hyper}, also uses S2 for geospatial processing.
PostGIS\footnote{\url{http://postgis.refractions.net/}}, a geospatial extension to PostgreSQL\footnote{\url{http://postgresql.org/}}, uses GEOS for geospatial processing and an R-tree implemented on top of GiST~\cite{hellerstein1995generalized} for indexing geospatial objects.
MonetDB, a column-oriented database, also uses GEOS for geospatial processing\footnote{\url{http://monetdb.org/Documentation/Extensions/GIS}}.
Magellan, a library for geospatial analytics that is based on Apache Spark, indexes
geospatial objects using the Z curve.
Similar to our technique, it is capable of identifying join partners in the
filter phase.
However, it uses a uniform grid and thus cannot adapt to the geometrical
features of the polygons and the expected distribution of query points.
Two popular Hadoop-based geospatial information systems are
Hadoop-GIS~\cite{aji2013hadoop} and SpatialHadoop~\cite{eldawy2014spatialhadoop}.
Both systems partition the data and store it in HDFS blocks such that spatial
objects that belong to a particular partition are stored in the same HDFS block
and insert the MBRs of the partitions into a global index.
Since these systems rely on offline partitioning of the data, they cannot
efficiently handle the online case where points are streamed.
GeoSpark~\cite{GeoSpark2015} does not support the online case either since it
does not allow to perform the join without an index on points, and
Simba~\cite{Simba2016} does not yet support polygons.
Most work on hardware optimizations for geospatial joins focuses on GPU
offloading\cite{BandiSAA2007SpatialJoinUsingGpu,YouZG2013GpuRtree,ZhangYG2017JoinProcessingOnGPUs}
and \cite{chavan2016towards} propose a vision for a GPU-accelerated end-to-end system
for performing spatial computations which decides whether to use CPUs or GPUs
based on a cost model.

\section{Conclusions}\label{sec:conclusions}
We have presented an adaptive geospatial join that leverages true hit
filtering using a hierarchical grid represented by a specialized radix tree.
We have transformed a traditionally compute-intensive problem into a
memory-intensive one.
Our approach is enabled by adaptively using the large high-bandwidth main
memory of modern hardware.
We have shown that it is possible to refine the index up to a user-defined
precision and identify all join partners in the filter phase.
We have demonstrated that the exact version of our algorithm can adapt to the
expected point distribution.
We have optimized our implementation for the brand new KNL processor and have
shown that our approach can outperform existing techniques by up to two orders
of magnitude.

In future work, we plan to evaluate our approach on the next-generation Intel
Xeon processor, named Skylake, which also supports AVX-512, and on GPUs.
We also plan to address offline workloads by leveraging small materialized
aggregates (SMAs) available in Data Blocks~\cite{datablocks}.

\section*{Acknowledgment}
We want to thank Ram Sriharsha for his help in evaluating Magellan.
This work has been sponsored by the German Federal Ministry of Education and Research (BMBF) grant 01IS12057 (FASTDATA and MIRIN).
This work is further part of the TUM Living Lab Connected Mobility
(TUM LLCM) project and has been funded by the Bavarian
Ministry of Economic Affairs and Media, Energy and Technology
(StMWi) through the Center Digitisation.Bavaria, an
initiative of the Bavarian State Government.
\balance

\bibliographystyle{IEEEtran}
\bibliography{geojoin}

\end{document}